\documentclass[12pt,english]{revtex4-1}
\usepackage{mathptmx}
\usepackage[T1]{fontenc}
\usepackage[latin9]{inputenc}
\usepackage{xcolor}
\usepackage{amsmath}
\usepackage{graphicx}
\PassOptionsToPackage{normalem}{ulem}
\usepackage{ulem}

\makeatletter

\providecolor{lyxadded}{rgb}{0,0,1}
\providecolor{lyxdeleted}{rgb}{1,0,0}
\DeclareRobustCommand{\lyxsout}[1]{\ifx\\#1\else\sout{#1}\fi}

\usepackage{chngcntr}
\counterwithout{equation}{section}
\counterwithout{figure}{section}

\usepackage{babel}

\makeatother

\usepackage{babel}
\begin{document}

\title{Competing exchange and irreversible reactions in a linear co-polycondensation
lead to a broad composition window where tunable high molecular weight
polymers can be prepared}

\author{Michael Lang$^{1*}$ and Frank Böhme$^{1}$}

\address{$^{1}$Leibniz-Institut für Polymerforschung Dresen, Hohe Straße
6, 01069 Dresden, Germany}
\email{lang@ipfdd.de}

\begin{abstract}
A co-polycondensation reaction is discussed analytically and by Monte-Carlo
simulations where two reactive units compete for reactions with an
alternating third reactive unit, whereby irreversible reactions replace
bonds which are able to undergo exchange reactions. The resulting
number average molar mass, $M_{\text{n}}$, exhibits only one distinct
peak at the stoichiometric condition of both competitors with the
alternating partner. The weight average molar mass, $M_{\text{w}}$,
reaches an additional second peak at the stoichiometric condition
between the dominating competitor and the alternating partner. Both
peaks of $M_{\text{w}}$ surround a range of compositions where a
rather high and approximately constant $M_{\text{w}}$ is obtained.
The degree of polymerization of the dominating and alternating reaction
partners is rather insensitive towards composition fluctuations if
the reaction mixture remains within this composition window. This
promotes high molecular weight species and more homogeneous weight
distributions at incomplete mixing conditions. An ideal reference
case (identical reaction rates for all reactions) is solved analytically
to describe these reactions. The position of the stable composition
window and the average molar masses inside this window can be tuned
by choosing appropriate precursor molecules, reaction mixtures or
post-tuning steps at later times. 
\end{abstract}
\maketitle

\section{Introduction}

Polycondensation is one of the basic classes of polymerization reactions
that is typically covered in textbooks of polymer science \citep{Flory,Rubinstein}.
The basic principle of condensation reactions has been understood
since the first half of the last century \citep{Kricheldorf}. Later,
developments focused mainly on the treatment of deviations from Flory's
ideality assumptions, which allowed to treat cyclization, substitution
effect, or reactions in open systems \citep{Kuchanov}. Despite of
the huge success made over the decades, the enormous amount of different
reaction pathways, boundary conditions of the reactions, etc ... still
allows for discoveries both on experimental and theoretical side.

Current work at our institute \citep{Suckow} on the co-polycondensation
(chain extension) of bisamino-terminated polyamide 12 (DD) with bisdithiooxalate
(AB-CC-BA), see Figure 1, provides an example for such a particular
situation. This co-polycondensation in the melt shows an extraordinary
increase in molar mass over an unexpectedly wide range of compositions.
This is largely surprising, as the key signature of a linear alternating
co-polycondensation at complete reactions is the formation of a narrow
peak in the average molar mass as a function of composition, see for
instance equation (\ref{eq:N_n}) and (\ref{eq:N_w1}) of the Appendix.
In practice, cyclization reduces both the number and weight average
degrees of polymerization \citep{Fischer,Kricheldorf2} and perfect
mixing conditions becomes crucial for a stoichiometric ratio $r\approx1$,
since the average molar mass diverges there. Also, impurities or inactive
groups may terminate a significant portion of the polymers. Most of
these deviations from ideal reactions were worked out in numerous
works that are summarized to some extent in Refs. \citep{Kricheldorf,Kuchanov}.
The general trend of these deviations is to cause a systematically
lower degree of polymerization in the vicinity of $r\approx1$. Therefore,
these deviations cannot explain an extended range of compositions
where high molecular mass products are available.

\begin{figure}
\begin{centering}
\includegraphics[width=1\columnwidth]{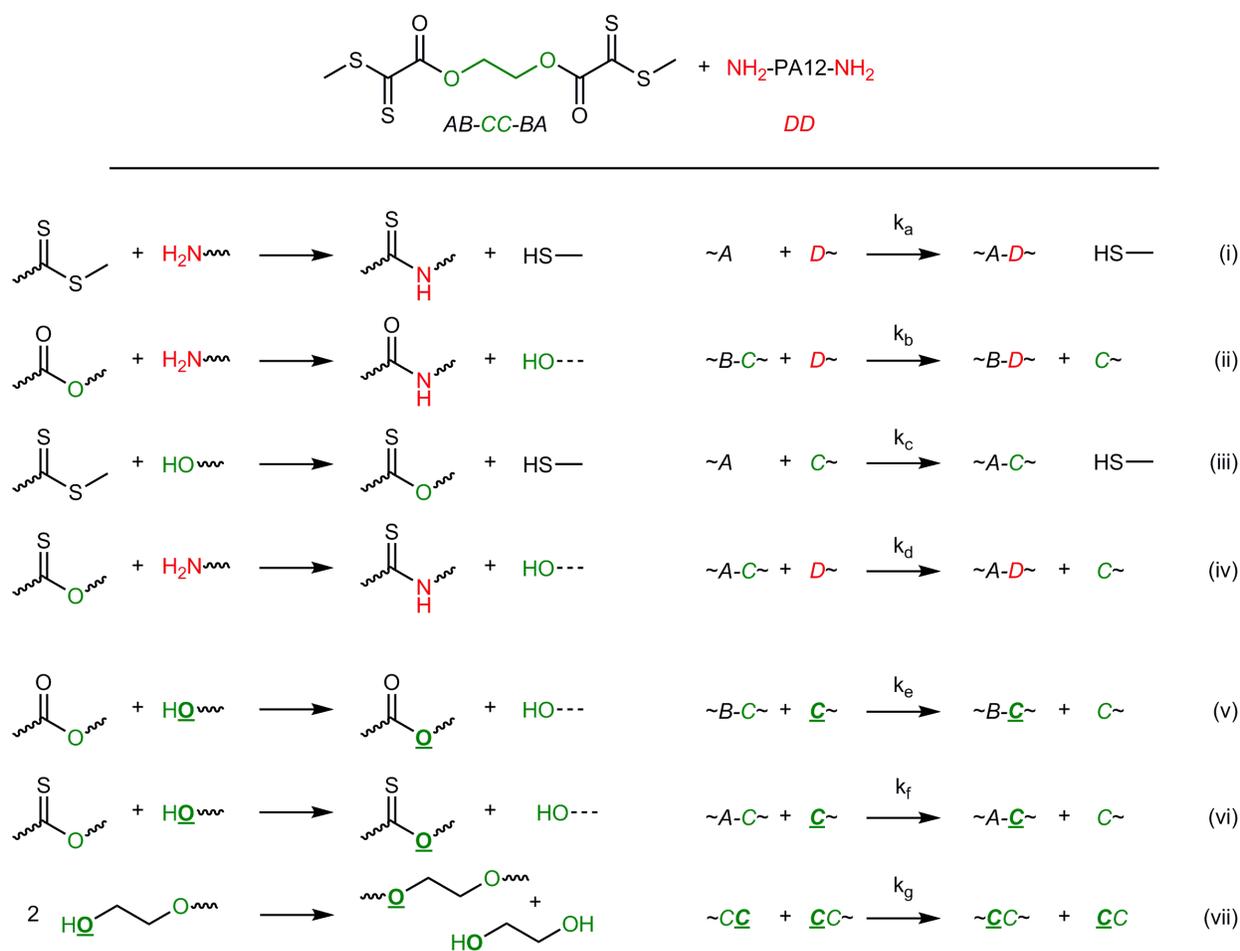}
\par\end{centering}
\caption{\label{fig:Scheme of reactions}Top: bisdithiooxalate and amino terminated
polyamide 12. Bottom: scheme of possible reactions. Taken from ref.
\citep{Suckow}.}
\end{figure}

A closer inspection of the possible reactions in Ref. \citep{Suckow}
indicates that the bisdithiooxalate contains four reactive centers
that potentially react with the amino groups at the ends of the polyamide
12, see Figure \ref{fig:Scheme of reactions}. Reactions with the
inner two of these reactive centers (B-C) result in the formation
of an amide group and a hydroxy group. The two dithioester groups
(A) at the ends of the bisdithiooxalate react with the amino or the
hydroxy groups. These reactions form either a thioamide group (A-D)
or a thioester group (A-C). Since reactions with the inner reactive
centers cut the bisdithiooxalate in two parts, only linear molecules
result from these reactions.

With this reaction scheme in mind, one is prompted directly to generalizations
of the alternating polycondensation reaction scheme where two different
reaction partners CC and DD compete for reactions with a third reaction
partner AB but do not react with one another. The expected behavior
of such a reaction is quite obvious: as long as the AB moieties are
in the majority with respect to the sum of CC and DD moieties, all
CC and DD moieties will be incorporated into the chains and reaction
rates affect only the sequence of CC and DD between the AB moieties.
If the AB moieties become minority, the reaction rates will control
how many CC and DD moieties can be found inside the chains that contain
AB moieties and how many of these remain non-reacted.

Such a line of arguments could be used for irreversible reactions
and a numerical solution can be obtained by integrating over the reaction
rates or running rate dependent Monte Carlo simulations, see section
\ref{sec:Kinetic-Model} and \ref{sec:Monte-Carlo-Simulations} below.
However, we have to emphasize that the amide and the thioamide groups
are stable, while the thioester and ester groups may undergo further
reactions with hydroxy or amino groups. In effect, the amino group
can displace the bound hydroxy groups. The hydroxy groups themselves
undergo an alcoholysis reaction that leads to a bond exchange similar
to a transesterification, see reactions (v)-(vii) in Figure \ref{fig:Scheme of reactions}.
Thus, from a general point of view, we discuss two kinds of competing
reactions, one is irreversible and the second one is an exchange reaction
that forms a special kind of a ``reversible'' bond where the density
of bonds is controlled by stoichiometry. We argue that this situation
refers to the competition of moieties CC and DD for bonds with AB,
where DD dominates over the competitor CC in such a way that the CC
moieties effectively react merely with the ``leftover'' reactive
groups after the coupling of AB with DD. This point is demonstrated
in our work by the excellent agreement of Monte-Carlo simulation that
reproduces the full reaction scheme of Figure \ref{fig:Scheme of reactions}
on one side, with analytical computations on the other side that consider
such a dominance of DD moieties over CC moieties.

The formation of reversible bonds or ``supramolecular polymerization''
\citep{DeGreef,Schmid} itself is well understood, in particular for
equilibrium systems \citep{Jacobson,Cates,Wittmer,Milchev}. Indeed,
even the competition between cyclization and linear polymerization
is solved (up to some minor details) for these systems in contrast
to irreversible polymerizations, where the history of reactions impacts
the resulting weight distributions \citep{Kricheldorf,Fischer}. However,
despite the large body of work on reversible and irreversible polycondensation
in general (see, for instance, Refs \citep{Kricheldorf,Kuchanov}
for a review), the authors could not find examples of competing reversible
and irreversible (or exchange) reactions within a linear condensation
co-polycondensation. Also, specialized works where CC and DD react
with AB \citep{Johnson}, publications on chain extension \citep{Yan,Yan2,Chalamet,Chalamet2},
exchange reactions (transesterfication, acidolysis, alcoholysis, ...)
\citep{Tobita,Qin,Iedema}, and competing irreversible reactions \citep{Skrobogatov},
or simulations works that model forward and backward reactions \citep{Krajniak}
did not provide models that may enable a quantitative discussion of
the experimental data of Ref. \citep{Suckow}.

This gap is closed with the present work. We first introduce a mean
field kinetic model that describes all possible reactions of Figure
\ref{fig:Scheme of reactions} in section \ref{sec:Kinetic-Model}.
Afterwards, we develop a Monte-Carlo model that resembles the reactions
of the kinetic model in order to capture the main aspects of the polycondensation
reaction in section \ref{sec:Monte-Carlo-Simulations}. These simulations
indicate that the reaction products are not much sensitive to reaction
rates for our model case as discussed in section \ref{sec:Dependence-on-reaction}.
Therefore, statistical arguments are used in section \ref{sec:Analytical-solution-of}
for the ``equireactive'' case (all reaction rates are identical)
to derive analytical expressions for the number and weight average
molar mass of the reaction products. The Monte-Carlo simulations allow
to compute the molar mass distribution of the full sample \citep{Bain,Brandao}
and specific weight distributions of particular moieties can be analyzed.
We use these to derive moiety specific degrees of polymerization as
molecules with or without a particular type of moiety may be separated
out of the reaction bath or may dominate the molar mass as in Ref.
\citep{Suckow}. One particularly interesting point of our results
is that a ``dominated co-polycondensation'' provides a range of
compositions, where the average molar mass of some reaction products
is rather independent of composition. How this composition range can
be tuned and how average molar masses or specific degrees of polymerization
can be adjusted ex posteriori is discussed in section \ref{sec:Discussion}.

\section{Kinetic Model\label{sec:Kinetic-Model}}

The kinetic equations that describe the reaction scheme of Figure
\ref{fig:Scheme of reactions} can be written as follows for all reactive
groups A, C, and D and all bonds A-D, A-C, B-C, and B-D, whereby bonds
containing C groups may undergo further reactions:

\begin{equation}
\frac{\text{d}\left[\text{A}\right]}{\text{d}t}=-k_{\text{a}}\left[\text{A}\right]\left[\text{D}\right]-k_{\text{c}}\left[\text{A}\right]\left[\text{C}\right]\label{eq:d1}
\end{equation}

\begin{equation}
\frac{\text{d}\left[\text{C}\right]}{\text{d}t}=k_{\text{b}}\left[\text{B-C}\right]\left[D\right]+k_{\text{d}}\left[\text{A-C}\right]\left[D\right]-k_{\text{c}}\left[\text{A}\right]\left[\text{C}\right]\label{eq:d2}
\end{equation}

\begin{equation}
\frac{\text{d}\left[\text{D}\right]}{\text{d}t}=-k_{\text{a}}\left[\text{A}\right]\left[\text{D}\right]-k_{\text{b}}\left[\text{B-C}\right]\left[\text{D}\right]-k_{\text{d}}\left[\text{A-C}\right]\left[\text{D}\right]\label{eq:d3}
\end{equation}
\begin{equation}
\frac{\text{d}\left[\text{A-C}\right]}{\text{d}t}=k_{\text{c}}\left[\text{A}\right]\left[\text{C}\right]-k_{\text{d}}\left[\text{A-C}\right]\left[\text{D}\right]\label{eq:d5}
\end{equation}

\begin{equation}
\frac{\text{d}\left[\text{A-D}\right]}{\text{d}t}=k_{\text{a}}\left[\text{A}\right]\left[\text{D}\right]+k_{\text{d}}\left[\text{A-C}\right]\left[\text{D}\right]\label{eq:d6}
\end{equation}
\begin{equation}
\frac{\text{d}\left[\text{B-C}\right]}{\text{d}t}=-k_{\text{b}}\left[\text{B-C}\right]\left[\text{D}\right]\label{eq:d7}
\end{equation}
\begin{equation}
\frac{\text{d}\left[\text{B-D}\right]}{\text{d}t}=k_{\text{b}}\left[\text{B-C}\right]\left[\text{D}\right].\label{eq:d8}
\end{equation}
The equations above reflect only reactions (i-iv) of the reaction
scheme, Figure \ref{fig:Scheme of reactions}, which change the concentrations
of the reactive groups or bonds. The alcoholysis reaction does not
affect the concentration of bonds or reactive groups, but rearranges
structure as indicated by reactions (v)-(vii) in Figure \ref{fig:Scheme of reactions}.
These need to be considered additionally for the Monte-Carlo simulations.

We further denote by $[\text{P}]_{0}$ the initial concentration of
the bisdithiooxalate molecules. As stoichiometric ratio $r$ we define
the ratio of reactive D groups with respect to the reactive groups
on AB moieties, 
\begin{equation}
r=\frac{[\text{D}]}{4[\text{P}]_{0}}=\frac{[\text{D}]}{[\text{A}]+[\text{B}]}.\label{eq:r-1}
\end{equation}

The coupled set of differential equations, equation (\ref{eq:d1})-(\ref{eq:d8})
was integrated numerically. The resulting concentrations of all reactive
groups and bonds were used to double check the corresponding results
of the Monte-Carlo simulations. We obtain that the final conversion
of A, C or D depends solely on the stoichiometric ratio $r$ and not
on reaction rates: 
\begin{itemize}
\item For $r<1/2$, a portion of $1-2r$ of the A groups remain non-reacted,
while all C and D groups are bound. 
\item For $1/2<r<1$, a portion of $2r-1$ of the C groups are not bound
to other groups, while all A and D groups are bound. 
\item For $r>1$, a portion of $1-1/r$ of the D groups is not reacted,
while all A groups are bound and all C moieties are not bound and
belong to non-reacted moieties CC. 
\end{itemize}
In consequence, the number average degree of polymerization, $N_{\text{n}}$,
depends solely on $r$ and not on reaction rates as $N_{\text{n}}$
is determined only by the concentration of chain ends (unbound A,
C, and D groups). However, such arguments do not apply for higher
order averages like weight averages or for specific degrees of polymerization
where averages over one type of moiety are performed, as these depend
on the corresponding weight distributions and thus, must depend on
reaction rates. This can be seen also from the numerical solution
of the rate equations, as the frequency of bonds that are formed is
indeed a function of the reaction rates.

Explicit variation of individual reaction rates by one order of magnitude
up or down shows only a weak impact on $N_{\text{w}}$ as discussed
in section \ref{sec:Dependence-on-reaction}. As it will turn out
below, a dependence on reaction rates is only possible when moieties
with different reactive groups on both ends are involved or initially
some groups are alreday reacted, which is not the most typical situation
for chain extension. Therefore, we focus in the following sections
on the case of an equal reactivity of all groups (as in similar work,
e.g. \citep{Macosko,Stockmayer}), moieties with identical reactive
groups on both ends, and initially unconnected moieties to derive
a simple analytical approximation. Similar to these works, we ignore
also the formation of cyclic molecules. This leads to a slight overestimation
of the molar mass \citep{Fischer} as most cyclic species are rather
short and consist of a stoichiometric amount of moieties for our alternating
co-polycondensation. This leads in effect to a higher density of the
majority units (i. e. chain ends) on the linear species, which reduces
additionally the average molar mass of the linear chains. 

\section{Monte-Carlo Simulations\label{sec:Monte-Carlo-Simulations}}

The bisdithiooxalate molecules are modeled by a sequence of three
moieties AB-CC-BA connected by two B-C bonds, see top of Figure \ref{fig:Scheme of reactions}.
DD moieties are modeled by single unconnected moiety at the beginning
of the simulations. A large number of $Z$ bisdithiooxalate molecules,
$10^{5}\le Z\le10^{6}$, are reacted with an amount $rZ$ of polyamide
12 within a Monte-Carlo scheme that resembles the reaction scheme
of Figure \ref{fig:Scheme of reactions}. All non-reacted A, C, and
D groups and all B-C and A-C bonds are considered as reactive units,
see section \ref{sec:Kinetic-Model}. Reaction rates $k_{\text{i}}$,
$i=a,...,d$ were normalized with respect to the largest reaction
rate $k_{\text{max}}$ among these, such that we can define acceptance
rates $p_{\text{i}}$ for particular reactions through 
\begin{equation}
p_{\text{i}}=\frac{k_{\text{i}}}{k_{\text{max}}},\label{eq:pi}
\end{equation}
where the $i=a,...,g$ indicates the reaction rates according to the
reaction scheme in Figure \ref{fig:Scheme of reactions}.

The Monte-Carlo scheme runs as follows: 
\begin{enumerate}
\item Select randomly a pair of reactive groups. 
\item Check whether these pair of groups can undergo a reaction (C or D
must react with A, A-C, or B-C). If not, return to step 1. 
\item Accept possible reactions with probability $p_{\text{i}}$. If rejected,
return to step 1. 
\item Perform reaction and update connectivity table and list of reactive
groups. 
\item Repeat steps 1-4 until either no more groups are reactive on DD moieties
(for $r<1$), or all A and B groups are reacted with D groups (for
$r>1$). 
\item For $r<1$: further equilibrate structure by additional reactions
of type (v)-(vii) in Figure \ref{fig:Scheme of reactions} such that
at least 5 interchange reactions were performed per reactive C group. 
\end{enumerate}
The first two steps assure that reactions occur proportional to the
concentrations of reactive groups. Effectively, a random selection
of reaction partners suppresses also cycle formation because of the
large numbers of molecules used. For our simulation parameters with
a number of $Z_{0}\ge10^{5}$ molecules at the start and $Z_{1}\ge10^{3}$
at the end (we do not approach critical points closer than 1\% of
conversion), the expected total number of rings formed is estimated
by summation over all ring closure probabilities, $\sum_{i=Z_{1}}^{Z_{0}}1/(2i-1)\le2.4$.
The resulting number fraction $<0.0024$ is too small to cause visible
deviations between simulation data and theoretical predictions. Nevertheless,
we also detected size of rings and considered them as linear chains
of same weight for the numerical analysis to further reduce the difference
between theory and simulation data.

Step 3 of the above scheme controls that the frequency of reactions
occurs with correct relative rates. For step 4 one has to be aware
that any reaction with an existing bond transfers the reactivity label
to the previously bound C group. Steps 5 and 6 control the end of
the reactions and allow for possible structural rearrangements (for
$r<1$) to drive the sample to equilibrium.

The above simulation scheme provides connectivity tables of all molecules.
These connectivity tables were used to determine the number and weight
average degrees of polymerization with respect to either all moieties
or a selected type of moiety. The results of this analysis for the
equireactive case are used as a reference for our analytical discussion
in the following sections. We also ran simulations with different
reaction rates to get an insight into the impact of different reaction
rates on the average molar mass. These results are discussed in section
\ref{sec:Dependence-on-reaction}.

\section{Analytical Solution of the Equireactive Case\label{sec:Analytical-solution-of}}

\subsection{General Remarks}

In the Appendix, we include a derivation of average and specific degree
of polymerization and the weight distributions for the simplest case
of an ideal alternating polycondensation of two monomers. This case
is included as a reference system and the scheme of derivation there
serves as a blueprint for the more complex situation of Figure \ref{fig:Scheme of reactions}
discussed below.

We have to point out that the initial condition of the reaction mixture
of ref. \citep{Suckow} introduces a non-random sequence of moieties,
since initially, all C groups are attached to B groups. Alcoholysis
helps to approach the random limit, as the bonds containing a C group
are continously regrouped. But the bonds with D groups are stable
and thus, still ``memorize'' the conditions during bond formation.
In fact, the initial condition of 100\% of C-B and no A-C bonds refers
to a maximum possible deviation between experimental data and the
analytical solution that is derived for the simplest ``equireactive
case'' where all reaction rates of Figure \ref{fig:Scheme of reactions}
are identical. As we shall see below, these differences are not very
large, which provides a strong argument in favor of using our analytical
solution as a reasonable approximation.

Below, we simplify the declaration of the structural units in the
top part of Figure \ref{fig:Scheme of reactions} and consider only
two-functional moieties of type A (the alternating partner), C (the
competitor), and D (the dominator). Thus, the equivalent initial set
up to Ref. \citep{Suckow} consists of a mixture of three connected
A-C-A moieties and isolated D moieties. We further generalize our
discussions by allowing for different ratios $s$ of C moieties with
respect to A moieties 
\begin{equation}
s=\frac{[C]}{[A]},\label{eq:s-2}
\end{equation}
while we keep the same variable $r$ for the ratio of reactive groups
on D moieties with respect to former AB (now A) moieties 
\begin{equation}
r=\frac{[D]}{[A]}.\label{eq:r-2}
\end{equation}
To simplify the derivation, we further introduce a portion $b$ of
bound groups of the species that terminates chains. In general, the
special solution for $s=1/2$ presented below refers to the situation
of Ref. \citep{Suckow}. However, additional corrections are necessary
for $r<1-s$ to approximate the effects that arise form the non-random
sequence of bonds that are caused by the specific initial conditions
used in Ref. \citep{Suckow}.

From a practical point of view, it might be interesting to know the
``specific degree of polymerization'' $z_{\text{X}}$ with $X=A,C,D$
that provides the average number of X moieties in chains that contain
at least one moiety X. Note that the reactions in the following sections
may form polymers where not all three types of moieties are present;
such polymers may phase separate from the reaction mixture or may
be taken out from the mixture by a chemistry specific separation technique
such as interaction chromatography. In the special case of Ref. \citep{Suckow},
there is one moiety much larger than the rest such the specific degree
of polymerization of this unit dominates the reaction mixture. In
such a case, the average molar mass is less suitable to characterize
the mixture \citep{Gao}. Note that we use here a symbol $z$ and
$z_{\text{X}}$ instead of $N$, or $N_{\text{X}}$ for the number
average degree of polymerization and the specific degrees of polymerization
with respect to moieties, since the moieties must not consist of single
monomers. This is also the case in Ref. \citep{Suckow} where the
moiety DD refers to polyamide 12.

Note that number and weight averages molar masses $M_{\text{n}}$
and $M_{\text{w}}$ might be available from Ref. \citep{Stockmayer,Macosko}
in some cases by a simplification of some more general results. However,
the specific average degrees of polymerization $z_{\text{X}}$ have
not been computed previously. Since one can derive $M_{\text{n}}$
and $M_{\text{w}}$ through $z_{\text{X}}$, we follow this route
and double check thereafter whether the average molar masses agree
with previous results, whenever these are available.

\subsection{\label{sec:D-terminated-chains:}D-terminated Chains: $[D]>[A]$}

We start with the simplest case of D-terminated chains, since here,
all C moieties are isolated at the end of the reactions, 
\begin{equation}
z_{\text{C}}=1,\label{eq:zcc-1}
\end{equation}
see section \ref{sec:Kinetic-Model}. Note that this refers in Ref.
\citep{Suckow} to the case of $r>1$. We start with considering an
ideal alternating polycondensation of A and D moieties. Here, the
D units are in the majority with respect to A units such that we introduce
a fraction of bound D groups (i.e. conversion of D groups) 
\begin{equation}
b=\frac{[A]}{[D]}=\frac{1}{r}\label{eq:s}
\end{equation}
to map the problem on the solution presented in the Appendix. Then,
$1-b$ is the probability that a given D group terminates a chain
made of A and D moieties or non-reacted D moieties. With the results
of the Appendix we obtain that both specific degrees of polymerization
are identical 
\begin{equation}
z_{\text{D}}=\frac{1}{1-b}=\frac{r}{r-1}=z_{\text{A}},\label{eq:zdd}
\end{equation}
even though each polymer contains one more D moieties than A moiety.
The reason for the above equivalence is that the non-reacted D moieties
do not contribute to $z_{\text{A}}$, see equation (\ref{eq:z_X})
of the Appendix. However, the non-reacted D moieties contribute to
the total number of moieties in AD polymers, $z_{\text{AD}}$. Since
there is always one A moiety less than D moieties for each chain,
we obtain 
\begin{equation}
z_{\text{AD}}=2z_{\text{D}}-1=\frac{1+b}{1-b}=\frac{r+1}{r-1},\label{eq:zdd-1}
\end{equation}
which is the classical result for a strictly alternating copolymerization,
see equation (\ref{eq:N_n}) of the Appendix.

Let us introduce $M_{\text{A}}$, $M_{\text{C}}$ and $M_{\text{D}}$
to denote the molar mass of an A, C, and D moiety respectively. A
portion of $1-b$ of all chains are non-reacted D moieties, while
the remaining chains have one more D moiety than A moieties. Thus,
the D containing chains have a number average molar mass of 
\[
M_{\text{n,AD}}=b\left(z_{\text{A}}M_{\text{A}}+\left(z_{\text{A}}+1\right)M_{\text{D}}\right)+\left(1-b\right)M_{\text{D}}=
\]
\begin{equation}
=\frac{M_{\text{A}}+rM_{\text{D}}}{r-1}.\label{eq:MnD}
\end{equation}

There are $r/\left(z_{\text{D}}s\right)$ chains with $z_{D}$ moieties
D per isolated moiety C with $z_{\text{C}}=1$ moieties C. This results
in an average number of moieties per chain of 
\begin{equation}
z=\left(\frac{r/\left(z_{\text{D}}s\right)}{r/\left(z_{\text{D}}s\right)+1}\right)z_{\text{AD}}+\frac{1}{r/\left(z_{\text{D}}s\right)+1}z_{\text{C}}\label{eq:z-1}
\end{equation}
\[
=\frac{s+r+1}{s+r-1}.
\]
The most interesting point of this result is that the average number
of moieties per molecule, $z$, diverges at a critical composition
of 
\begin{equation}
r_{\text{c}}=1-s,\label{eq:rc}
\end{equation}
which is at a different position than the critical points of the specific
degrees of polymerization $z_{\text{D}}$ and $z_{\text{A}}$, which
diverge at 
\begin{equation}
r_{\text{c}}=1.\label{eq:rc2}
\end{equation}
This is qualitatively different to an irreversible alternating co-polycondensation,
where all degrees of polymerization (specific \& average) diverge
at the same critical point (cf. with the Appendix).

\begin{figure}
\includegraphics[angle=270,width=1\columnwidth]{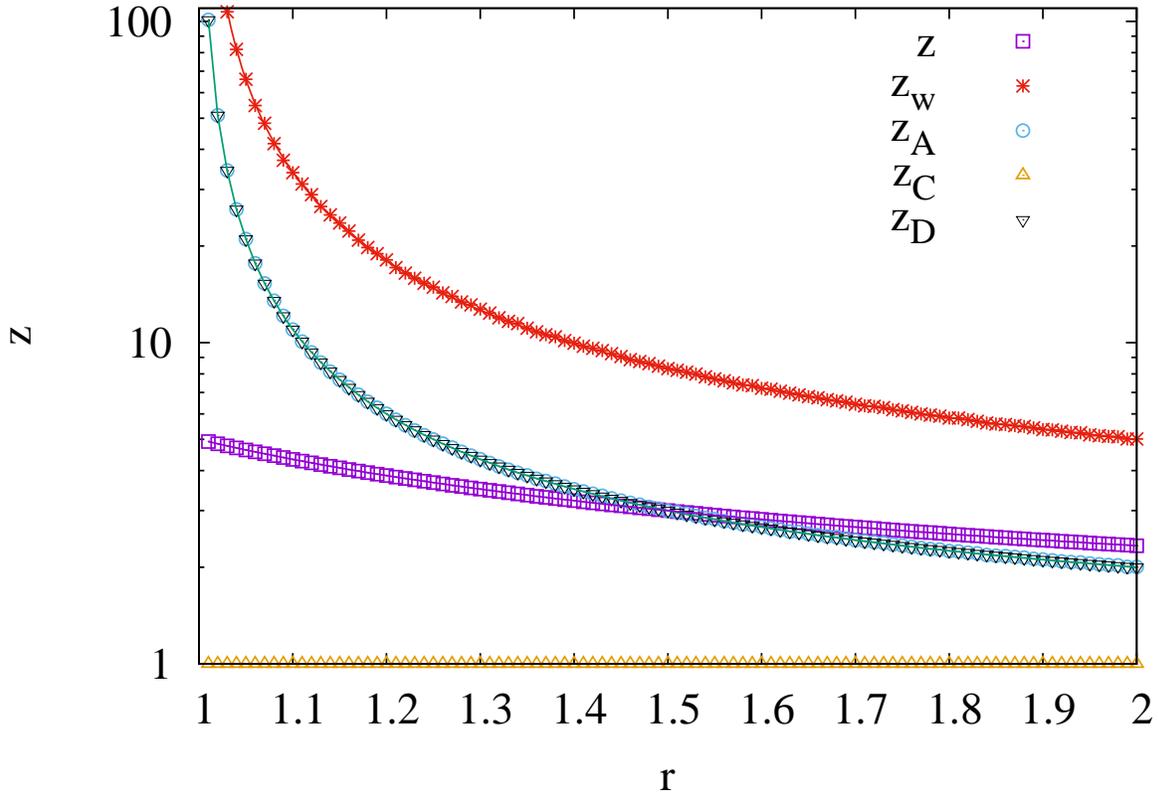}

\caption{\label{fig:Average-degrees-of}Average degrees of polymerization for
$r>1$ and $s=1/2$ (specific case of Ref. \citep{Suckow}). Comparison
of Monte-Carlo simulation data (data points) and the corresponding
analytical solutions for the equireactive case (all reaction rates
are identical) as given in the text and shown by the lines.}
\end{figure}

Let us introduce the composition average molar mass per moiety: 
\begin{equation}
\overline{M}=\frac{M_{\text{A}}+sM_{\text{C}}+rM_{\text{D}}}{r+s+1}.\label{eq:Mquer}
\end{equation}
With this definition, one obtains for the number average molar mass
of all molecules that 
\[
M_{\text{n}}=\frac{\left(r-1\right)M_{\text{n,AD}}}{r-1+s}+\frac{sM_{\text{C}}}{r-1+s}
\]
\begin{equation}
=\frac{M_{\text{A}}+sM_{\text{C}}+rM_{\text{D}}}{r-1+s}=z\overline{M}\label{eq:Mn}
\end{equation}
All results of this section are exact and confirmed by simulation
data, see Figure \ref{fig:Average-degrees-of}, where the predictions
above are double checked with Monte-Carlo simulation data.

The termination argument used to derive equation (\ref{eq:zdd}) indicates
that the weight distribution of $z_{\text{A}}$ and $z_{\text{D}}$
are most probable ones. The number fraction distribution, $n_{\text{N}}$,
of chains containing either $N$ moieties A or $N$ moieties D (among
all chains containing at least one A or at least one D moiety) are
identical and of most probable type: 
\begin{equation}
n_{\text{N,A}}=n_{\text{N,D}}=\left(1-b\right)b^{N-1}=\left(r-1\right)r^{-N}.\label{eq:nND}
\end{equation}
This leads to a polydispersity of \citep{Rubinstein} 
\begin{equation}
\frac{z_{\text{w,A}}}{z_{\text{A}}}=\frac{z_{\text{w,D}}}{z_{\text{D}}}=1+b=\frac{r+1}{r},\label{eq:pdi}
\end{equation}
since $b$ is the fraction of reacted groups that connect to another
molecule, i.e. it is equivalent to conversion for the classical case
of polycondensation of a single type of monomer. Therefore, 
\begin{equation}
z_{\text{w,A}}=z_{\text{w,D}}=\frac{1+b}{1-b}=\frac{r+1}{r-1}=z_{\text{AD}}.\label{eq:zwA}
\end{equation}

\begin{figure}
\includegraphics[angle=270,width=1\columnwidth]{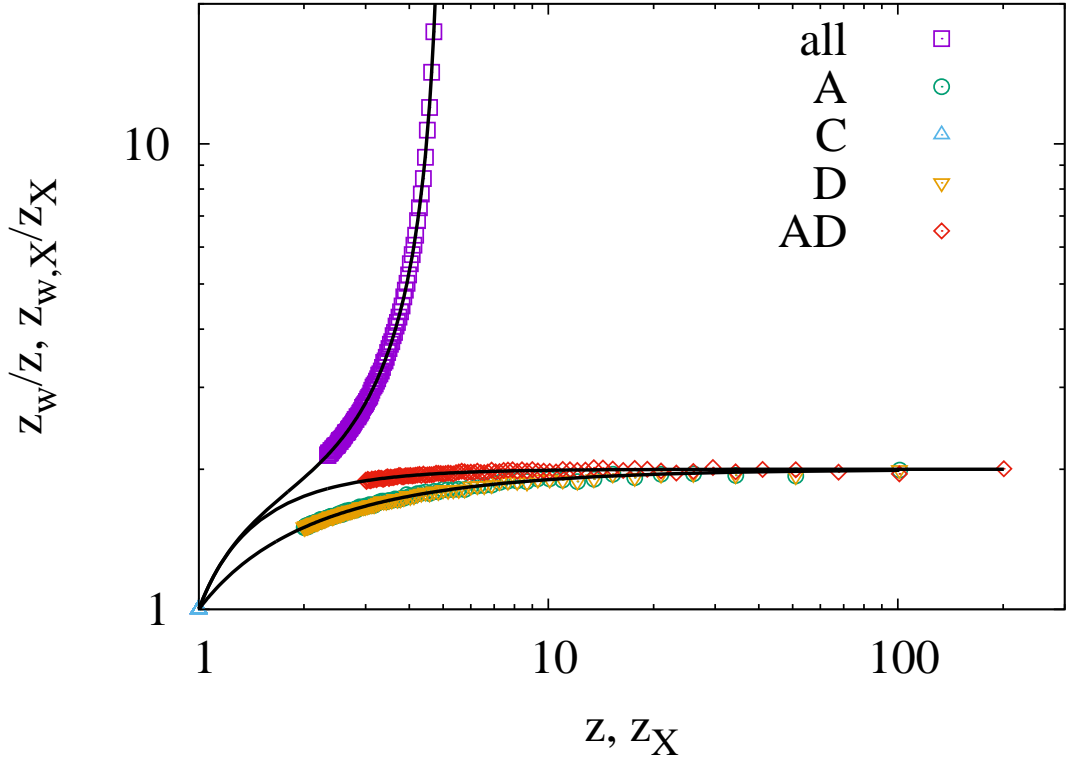}

\caption{\label{fig:Polydispersity-of-different}Polydispersity of different
molecular weights for $r>1$ for the special case $s=1/2$ of Ref.
\citep{Suckow}. Simulation data (symbols) is compared with the corresponding
theoretical predictions, equation (\ref{eq:pdi}), (\ref{eq:zwD}),
and (\ref{eq:zwz}).}
\end{figure}

The weight average molar mass of the whole sample is obtained by averaging
(according to the corresponding weight fractions) the weight averages
of independent distributions (distribution of C moieties is independent
of the combined AD distribution). The molar mass of a chain containing
$N-1$ moieties of type A and $N$ moieties of type D is 
\begin{equation}
M_{\text{N,AD}}=\left(N-1\right)M_{\text{A}}+NM_{\text{D}}.\label{eq:MND}
\end{equation}
The weight average molar mass of all D containing chains is then computed
using the first and the second moment, $m_{1}$ and $m_{2}$ of the
number fraction distribution 
\begin{equation}
M_{\text{w,AD}}=\frac{m_{2}}{m_{1}}=\frac{\sum_{N=1}^{\infty}n_{\text{N,AD}}M_{\text{N,AD}}^{2}}{M_{\text{n,AD}}}\label{eq:MwD}
\end{equation}

\[
=\frac{\left(b+b^{2}\right)M_{\text{A}}^{2}+\left(1+b\right)M_{\text{D}}^{2}+4bM_{\text{A}}M_{\text{D}}}{\left(1-b\right)\left(bM_{\text{A}}+M\right)}.
\]
\[
=\frac{\left(r+1\right)M_{\text{A}}^{2}+\left(r^{2}+r\right)M_{\text{D}}^{2}+4rM_{\text{A}}M_{\text{D}}}{\left(r-1\right)\left(M_{\text{A}}+rM_{\text{D}}\right)}.
\]
This result agrees with a simplified version of equation (3) of Ref.
\citep{Stockmayer} (see also corrections to Ref. \citep{Stockmayer}).
By setting $M_{\text{A}}=M_{\text{D}}=1$ we obtain the weight average
number of moieties in the D containing chains 
\begin{equation}
z_{\text{w,AD}}=\frac{b^{2}+6b+1}{1-b^{2}}=\frac{r^{2}+6r+1}{r^{2}-1}.\label{eq:zwD}
\end{equation}
Note that this result was obtained directly in the Appendix, equation
(\ref{eq:N_w1}), and is the classical result for a strictly alternating
linear polycondensation. Note that the corresponding polydispersity
approaches 2 for $r\rightarrow1$ similar to an ideal condensation
polymerization, but with a different function, see Figure \ref{fig:Polydispersity-of-different}
and equation (\ref{eq:PDI}).

In order to obtain $M_{\text{w}}$ and $z_{\text{w}}$ of the full
sample, we have to combine $M_{\text{w,AD}}$ with $M_{\text{\text{w},C}}=1$
with respect to their weight fractions within the sample. There is
a number fraction of $s/\left(1+r+s\right)$ C moieties in the sample.
Thus, the weight fraction 
\begin{equation}
w_{\text{C}}=\left(\frac{s}{1+r+s}\right)\frac{M_{\text{C}}}{\overline{M}}\label{eq:wC}
\end{equation}
of C moieties in the sample is combined with the above results for
$M_{\text{w,AD}}$ and $w_{\text{C}}$ to provide 
\begin{equation}
M_{\text{w}}=\left(1-w_{\text{C}}\right)M_{\text{w,AD}}+w_{\text{C}}M_{\text{C}}.\label{eq:Mw}
\end{equation}
For $M_{\text{A}}=M_{\text{C}}=M_{\text{D}}=1$, the resulting expression
can be simplified to

\begin{equation}
z_{\text{w}}=\frac{r^{2}+6r+sr-s+1}{\left(r-1\right)\left(1+r+s\right)}.\label{eq:zw}
\end{equation}
The corresponding polydispersity is here

\begin{equation}
\frac{z_{\text{w}}}{z}=\frac{\left(r^{2}+6r+sr-s+1\right)\left(r+s-1\right)}{\left(r-1\right)\left(1+r+s\right)^{2}}\label{eq:zwz}
\end{equation}
which diverges for $r\rightarrow1$, since $z_{\text{w}}$ diverges,
while $z$ remains finite (see also Figure \ref{fig:Average-degrees-of}).
This is a rather unusual behavior for a linear condensation polymerization,
where usually both averages diverge at the critical point while the
polydispersity remains finite. In our case, the divergence is caused
by the presence of a large number of C moieties that are expelled
from the chains and that keep $z$ finite. This situation is qualitatively
similar to a non-linear poly-condensation where the abundant sol molecules
prevent the number average molar mass from divergence at the gel point.
The main results of the analytical derivation above were double checked
with simulation data showing perfect match between both, see Figure
\ref{fig:Polydispersity-of-different}.

\subsection{A-terminated Chains: $[A]>[C]+[D]$\label{subsec:A-terminated-chains:}}

Since the number of A moieties is larger than the number of C and
D moieties for this case, all chain ends are established by A moieties,
see section \ref{sec:Kinetic-Model}. Thus, all C and D moieties are
completely reacted at full conversion and serve as bonds between A
moieties. Note that this case is equivalent to the condition $r+s<1$,
which refers for $s=1/2$ to the case $r<1/2$ in Ref. \citep{Suckow}.
For this limiting case, there are significant correlations due to
the special chemistry of Ref. \citep{Suckow}, which starts from a
non-random sequence of moieties where the A moieties have two different
reactive sites at their ends. This non-randomness affects higher order
averages of molar mass and some specific averages, but it must disappear
from the distribution in the limit $r\rightarrow1$ where all ``initial''
bonds with C moieties are replaced by new bonds with D moieties. Therefore,
we have to specify our results for the random case further to apply
these to the particular situation of Ref. \citep{Suckow}.

The most intuitive way to derive the general solution of the random
case is by using a probability $b$ that describes the density of
bound groups of the majority A moieties (i.e. the conversion of A
groups) 
\begin{equation}
b=\frac{[C]+[D]}{[A]}=s+r,\label{eq:q_new}
\end{equation}
and the mixture of C and D moieties by a probability $p$ that a ``bond''
between A moieties is a D moiety:

\begin{equation}
p=\frac{[D]}{[D]+[C]}=\frac{r}{r+s}.\label{eq:p}
\end{equation}
There are in average 
\begin{equation}
z_{\text{A}}=\frac{1}{1-b}=\frac{1}{1-s-r}\label{eq:z_AB}
\end{equation}
A moieties per chain. The number of ``bond'' moieties (C or D) is
always one less: $z_{A}-1$. Thus, the average number of moieties
per chain is 
\begin{equation}
z=2z_{\text{A}}-1=\frac{1+b}{1-b}=\frac{1+s+r}{1-s-r}.\label{eq:z}
\end{equation}

The average molar mass of a chain inside the reaction mixture is computed
by considering $p$ to split the portion of connecting moieties into
C and D type. Using the above results we obtain for the number average
molar mass, $M_{\text{n}}$, that 
\begin{equation}
M_{\text{n}}=z_{\text{A}}M_{\text{A}}+\left(z_{\text{A}}-1\right)\left(pM_{\text{D}}+\left(1-p\right)M_{\text{C}}\right)\label{eq:M_n}
\end{equation}
\[
=\frac{M_{\text{A}}+rM_{\text{D}}+sM_{\text{C}}}{1-s-r}=z\overline{M}.
\]

For deriving $z_{\text{D}}$ and $z_{\text{C}}$, we assume a random
sequence of C and D along all chains. For computing $z_{\text{D}}$,
we need to leave aside all chains that are entirely connected by C
units and vice versa. As in the above section, we make use of the
point that the number fraction distribution of a sequence of $N\ge1$
``bonds'' is equivalent to the most probable number fraction distribution,
$n_{\text{N}}$, of a chain containing $N$ units, which is terminated
with probability $1-b$ that a connection is missing, 

\begin{equation}
n_{\text{N}}=\left(1-b\right)b^{N-1}=\left(1-s-r\right)\left(s+r\right)^{N-1}.\label{eq:nN-1}
\end{equation}
The probability that a sequence of $N$ connecting groups is of type
C is $(1-p)^{N}$. The total number fraction of chains that contain
exactly $N$ moieties of type C (and thus no D moiety) is therefore
\[
n_{\text{N,C}}=\sum_{N=1}^{\infty}n_{\text{N}}(1-p)^{N}=\frac{1-b}{b}\sum_{N=1}^{\infty}\left(b\left(1-p\right)\right)^{N}
\]

\begin{equation}
=\frac{1-s-r}{s+r}\sum_{N=1}^{\infty}s^{N}=\frac{\left(1-s-r\right)s}{\left(s+r\right)\left(1-s\right)}.\label{eq:N_NC}
\end{equation}
Recall from the case $r>1$ that the specific degree of polymerization
of the minority species is the same as the specific degree of polymerization
of the majority moieties, see equation (\ref{eq:zdd}). Thus, the
average number of D moieties per all molecules is the average number
of C and D moieties per molecule times $p$, which gives $z_{\text{A}}p$.
These D moieties, however, are only located on the number fraction
of $1-n_{\text{NC}}$ chains that contain at least on D moiety. This
increases the average number of D moieties per D containing chain,
$z_{\text{D}}$, from $z_{\text{A}}p$ to 
\begin{equation}
z_{\text{D}}=\frac{z_{\text{A}}p}{1-n_{\text{N,C}}}=\frac{1-s}{1-s-r}\label{eq:zDD}
\end{equation}

\begin{figure}
\includegraphics[angle=270,width=1\columnwidth]{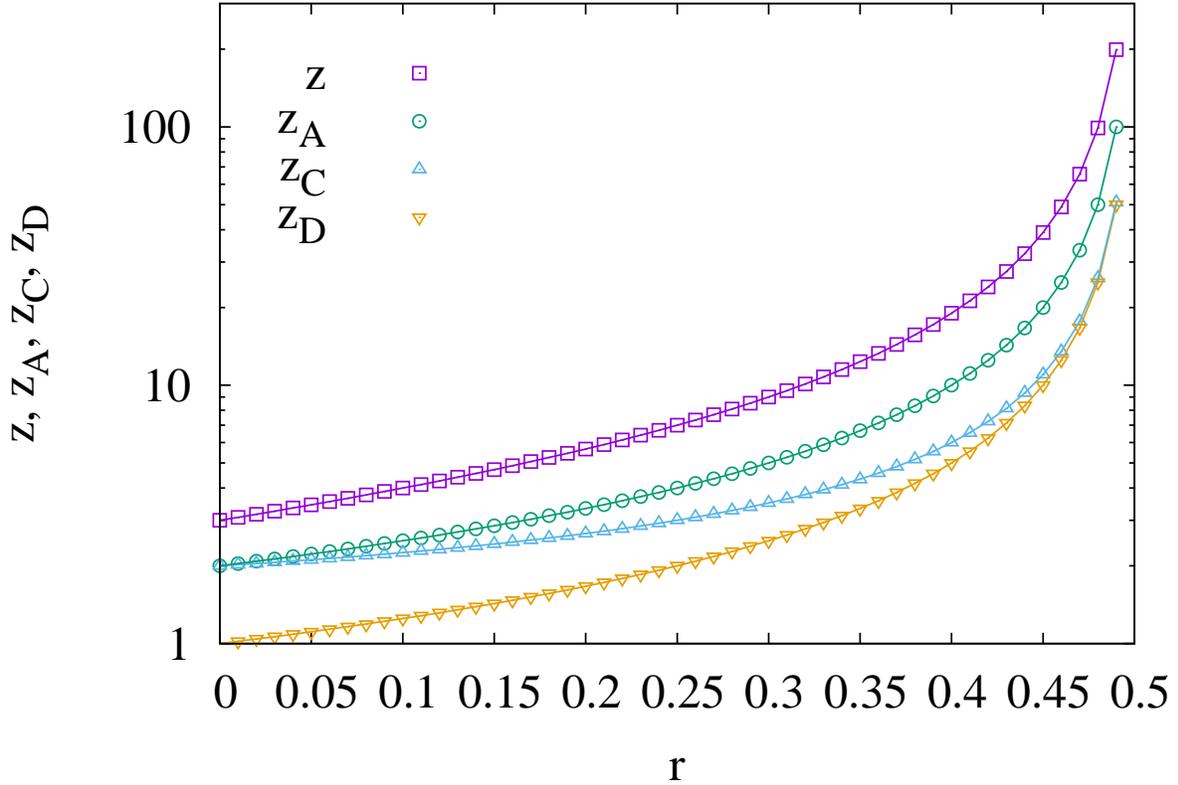}

\caption{\label{fig:Degrees-of-polymerization-1}Degrees of polymerization
from Monte-Carlo simulations (symbols) of the random case as compared
to the model predictions for $r<1-s$ with $s=1/2$, which are shown
with lines.}
\end{figure}

With similar arguments, one obtains the number fraction of bond chains
containing exactly N moieties D 
\begin{equation}
n_{\text{N,D}}=\frac{1-b}{b}\sum_{N=1}^{\infty}\left(pb\right)^{N}=\frac{\left(1-s-r\right)r}{\left(s+r\right)\left(1-r\right)},\label{eq:nND-2}
\end{equation}
which leads to 
\begin{equation}
z_{\text{C}}=\frac{z_{\text{A}}\left(1-p\right)}{1-n_{\text{N,D}}}=\frac{1-r}{1-r-s}.\label{eq:zc}
\end{equation}
As before, our results are checked against simulation data, here for
the equireactive random case (no initial B-C bonds), see Figure \ref{fig:Degrees-of-polymerization-1}.

In contrast to the random case, the situation in Ref. \citep{Suckow}
enforces $z_{\text{C}}=1$ at $r=0$ instead of $z_{\text{C}}=2$.
Furthermore, correlations among the sequence of C and D moieties are
introduced through the different ends of the A moieties. These survive
in part from the enforced initial conditions and are introduced to
some other part through different reaction rates of C and D with the
different ends of A. The third principal difference to the random
case is, that at $r=0$ despite of $s>0$ all A moieties are chain
ends.

As obvious from the above derivation, $M_{\text{n}}$, $z$, and $z_{\text{A}}$
depend only on the density of ends and not on the sequence of bonds
- in contrast to $z_{\text{C}}$ and $z_{\text{D}}$. Therefore, the
above results for $M_{\text{n}}$, $z$, and $z_{\text{A}}$ hold
also for the simulation data that resemble the situation of Ref. \citep{Suckow}.
A simple zero order approximation for $z_{\text{C}}$ and $z_{\text{D}}$
in Ref. \citep{Suckow} is to consider that the initial conditions
were equivalent to the polycondensation of A-C-A unit with D moieties
(assuming no exchange of bonds with C moieties). Then, simply, 
\begin{equation}
z_{\text{C}}=\frac{z_{\text{A}}}{2}=z_{\text{D}},\label{eq:zcc}
\end{equation}
whereby the last equivalence is inferred from the results for the
two partners of an ideal polycondensation (in the above simplification,
D and C moieties are alternating along the chains), see the Appendix.

\begin{figure}
\includegraphics[angle=270,width=1\columnwidth]{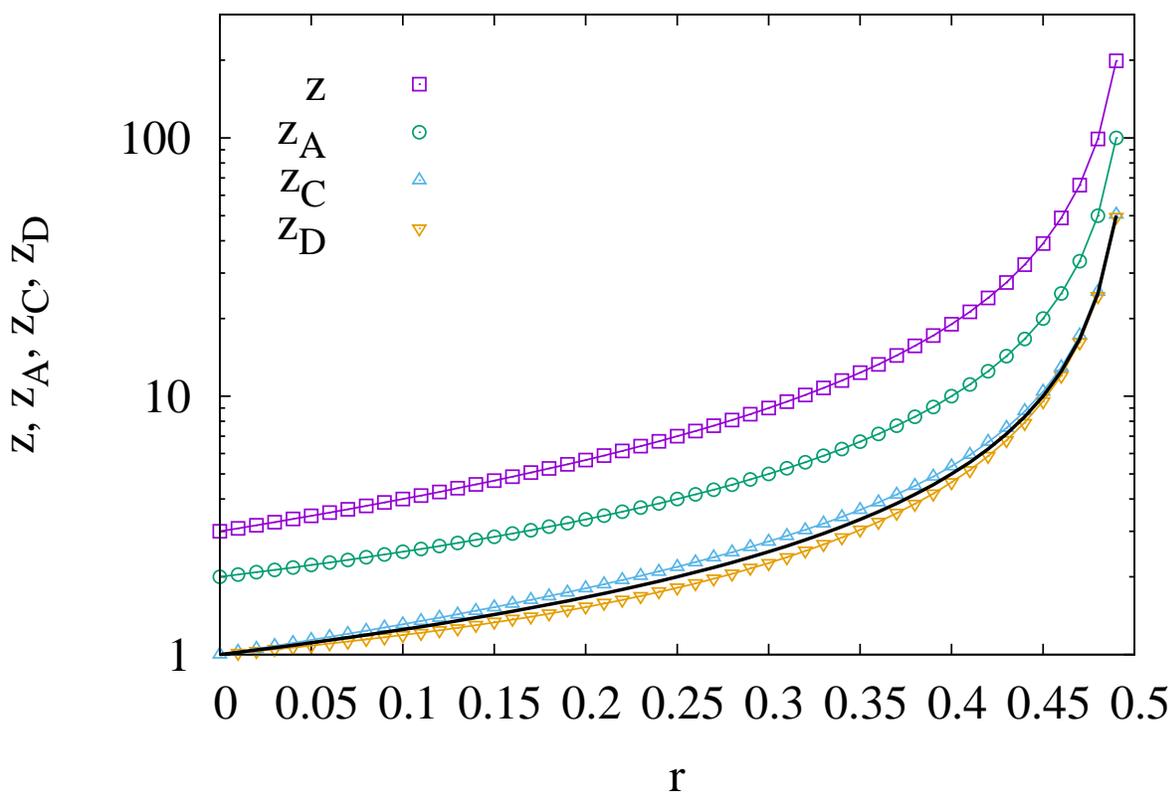}

\caption{\label{fig:Degrees-of-polymerization}Degrees of polymerization from
Monte-Carlo simulations (symbols) that resemble the special conditions
of Ref. \citep{Suckow} for $r<1-s$ with $s=1/2$ compared with theoretical
predictions. The thick black line indicates the simple approximation,
equation (\ref{eq:zcc}), while the lines for $z_{\text{C}}$ and
$z_{\text{D}}$ indicate the exact solutions mentioned below equation
(\ref{eq:zcc}).}
\end{figure}

A comparison with the Monte-Carlo simulation data in Figure \ref{fig:Degrees-of-polymerization}
shows perfect agreement of the theoretical predictions with $z$ and
$z_{\text{A}}$ of the majority species for simulations, that resemble
the equireactive case with initial conditions as in Ref. \citep{Suckow}
(A-C-A initial sequence of all A and C moieties). The data for the
minority species, $z_{\text{D}}$ and $z_{\text{C}}$, are close to
the simple approximation of equation (\ref{eq:zcc}). A more detailed
discussion of the correlations introduced by initial conditions and
propagated by the asymmetry of AB moieties (not included here) shows,
that the solutions for the special case of Ref. \citep{Suckow} are
shifted with respect to the simple approximation by $\pm\left(r/2+r^{2}\right)$
in the equireactive case, see also Figure \ref{fig:Degrees-of-polymerization}.
Since this shift is still in the range of a 10\% correction, we argue
that a simplifcation similar to equation (\ref{eq:zcc}) should provide
a reasonable approximation for $z_{\text{C}}$ and $z_{\text{D}}$
of the minority species in similar cases.

\begin{figure}
\includegraphics[angle=270,width=1\columnwidth]{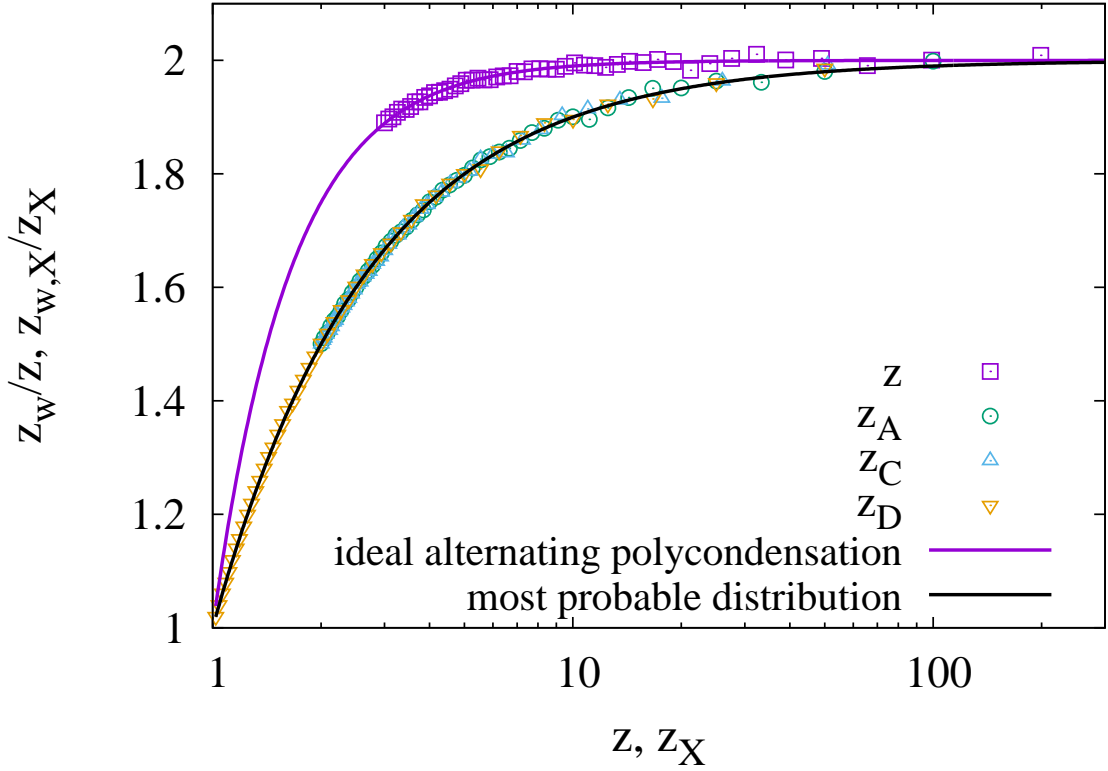}

\caption{\label{fig:Polydispersity-of-the-1}Polydispersity for $r<1-s$ and
$s=1/2$ plotted as a function of the average degree of polymerization.
While all specific degree of polymerization $z_{\text{X}}$, $X=A,C,D$
are of most probable type, see equations (\ref{eq:zwa})-(\ref{eq:zwd}),
$z$ is of an ideal alternating polycondensation type, see equation
(\ref{eq:zw-2}).}
\end{figure}

Let us now return to the random case of equally reactive groups. For
the derivation of the weight averages, we make use of the findings
for $r>1$ and of the derivation in the Appendix. First, all specific
``degrees of polymerization'' $z_{\text{X}}$ with $X=A,C,D$ are
characterized by a most probable number fraction distribution, see
the above derivation. Therefore, the polydispersity of these special
degrees of polymerization $z_{\text{X}}$ is the same as for a most
probable distribution as a function of the average degree of polymerization,
see also equation (\ref{eq:poly Z}) of Appendix. This provides directly
\begin{equation}
z_{\text{w,A}}=z_{\text{A}}\left(1+b\right)=\frac{1+s+r}{1-s-r}=z,\label{eq:zwa}
\end{equation}
\begin{equation}
z_{\text{w,C}}=z_{\text{C}}\left(1+b\right)=\left(1-r\right)z,\label{zwc}
\end{equation}
\begin{equation}
z_{\text{w,D}}=z_{\text{D}}\left(1+b\right)=\left(1-s\right)z.\label{eq:zwd}
\end{equation}
Similarly, the average number of moieties per chain must be characterized
by the polydispersity of an ideal co-polycondensation, since the A
moieties form a perfectly alternating sequence with the remaining
moieties. Using equation (\ref{eq:PDI}) of the Appendix we arrive
at 
\begin{equation}
z_{\text{w}}=z\frac{1+6b+b^{2}}{\left(1+b\right)^{2}}=\frac{1+6(r+s)+\left(r+s\right)^{2}}{\left(1-r-s\right)\left(1+r+s\right)}.\label{eq:zw-2}
\end{equation}
These results are tested by comparing the polydispersities of the
different cases with simulation data in Figure \ref{fig:Polydispersity-of-the-1}.

\begin{figure}
\includegraphics[angle=270,width=1\columnwidth]{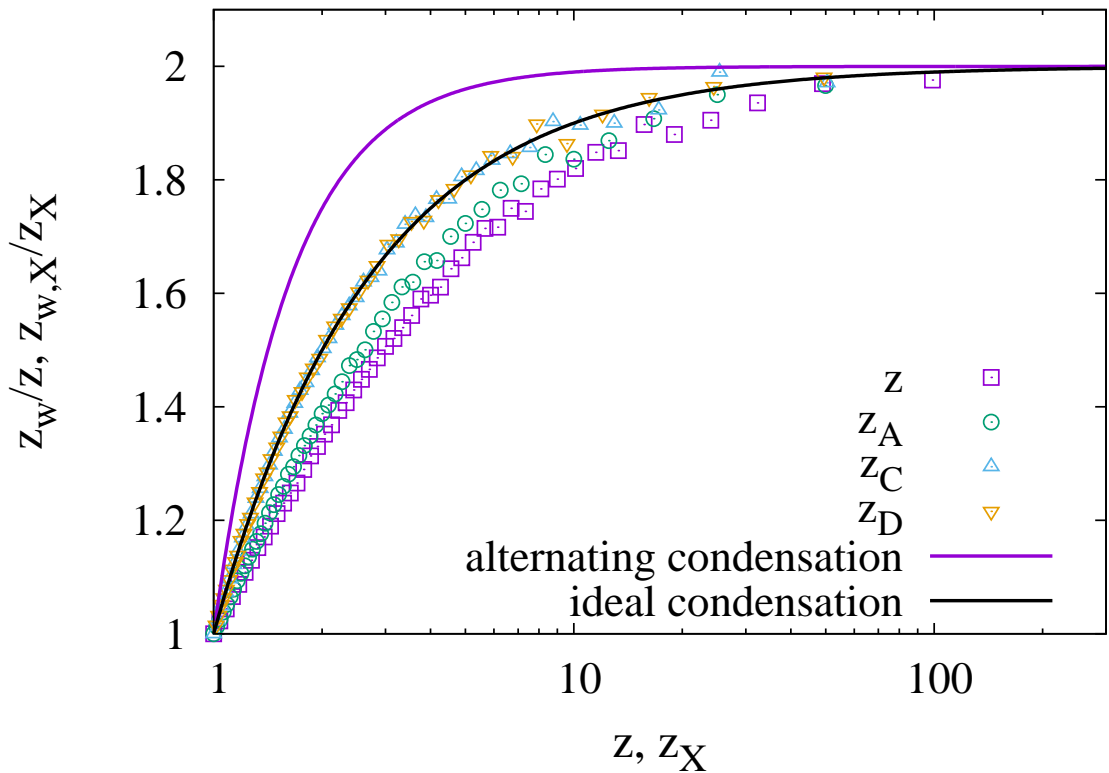}

\caption{\label{fig:Polydispersity-of-the}Polydispersity as a function of
$z$ (note the re-normalization of $z_{\text{A}}$ and $z$ discussed
around equation (\ref{eq:z_0})) for the specific situation of ref.
\citep{Suckow} with $s=1/2$. }
\end{figure}

For the special situation of Ref. \citep{Suckow}, we adopt the same
simple approximation as above and interpret the initial conditions
as a polycondensation of A-C-A unit with D moieties without regrouping
of bonds with C moieties. Thus, we expect that all specific degrees
of polymerization show the polydispersity of a most probable distribution.
The only necessary correction is to renormalise $z_{\text{A}}$ by
a factor of 1/2 to take into account that each A-C-A unit contributes
two A moieties. Similarly, we scale $z$ down by a factor of 
\begin{equation}
z_{0}=\frac{3+2r}{1+2r},\label{eq:z_0}
\end{equation}
which is the average degree of polymerization at the onset of the
reactions. Figure \ref{fig:Polydispersity-of-the} shows that our
expectation is fully satisfied for $z_{\text{C}}$ and $z_{\text{D}}$
while the simple re-scaling leads to less satisfactory results for
$z$ and $z_{\text{A}}$ where polydispersity is clearly overestimated.
Still, the error for $z_{A}$ is in the range of 10\%, while for the
special situation of Ref. \citep{Suckow} the polydispersity of a
polycondensation might be taken as a reference for the re-scaled $z_{\text{w}}$
instead of the expected co-polycondensation case. The reason for the
lower polydispersity is the suppression of species that contain only
single moieties at intermediate degrees of polymerization, which leads
to a consecutive suppression of the high molecular weight tail, as
$z$ is fixed by the density of ends (our simple re-scaling corrects
only initial conditions).

Therefore, we expect the data of the experiments of Ref. \citep{Suckow}
in the vicinity of 
\begin{equation}
z_{\text{w}}\approx z\left(1+b\right)=\frac{\left(1+r+s\right)^{2}}{1-r-s},\label{eq:zw-3}
\end{equation}
and similarly, we approximate 
\begin{equation}
M_{\text{w}}\approx M_{\text{n}}\left(1+b\right)\approx z_{\text{w}}\overline{M}.\label{eq:mw}
\end{equation}

\subsection{\label{sec:C-terminated-chains:}C-terminated Chains: $[D]<[A]<[C]+[D]$}

In this section, we combine the results of the preceding two sections,
as both critical points at the boundaries control the average degrees
of polymerization. A comparison of our simulation data for the random
case and the one with initial conditions as in Ref. \citep{Suckow}
shows that the correlations from initial conditions can be ignored
here. The physical explanation for this observation is that the C
moieties, which carry the correlation to the initial conditions, are
increasingly expelled from the chains with increasing $[D]$ such
that initial correlations or the effect of a different reactivity
become negligible for $r\rightarrow1$. Since the degrees of polymerizations
are typically $\gg1$ for this section, the remaining small correlations
are averaged out to a large extent. Therefore, we derive here the
solutions for the random case and compare directly with data for the
specific simulations that resemble the situation of Ref. \citep{Suckow}
for the case $1-s<r<1$. Agreement between both supports the proposal
that correlations can be ignored.

Let 
\begin{equation}
b=\frac{[A]-[D]}{[C]}=\frac{1-r}{s}\label{eq:q-1}
\end{equation}
denote the fraction of the bound C groups with respect to all C groups.
Since $1-b$ defines the probability that a C group terminates a chain
of C units (leaving aside the completely reacted A and D moieties
in between), we have 
\begin{equation}
z_{\text{C}}=\frac{1}{1-b}=\frac{s}{s+r-1}.\label{eq:zcc-2}
\end{equation}
Again, a fraction of $1-b$ of all chains are isolated C moieties.
Thus, all A moieties are distributed only among a portion $b$ of
chains. Since there are in average $1/s$ moieties A per C moiety,
we have 
\begin{equation}
z_{\text{A}}=\frac{z_{\text{C}}}{sb}=\frac{s}{\left(1-r\right)\left(s+r-1\right)}.\label{eq:z_AB-1}
\end{equation}

Similar to the preceding cases, the distribution of $z_{\text{A}}$
is of an ideal condensation polymerization type, however, there are
now two critical points that control $z_{\text{A}}$ as indicated
by the denominator of equation (\ref{eq:z_AB-1}). But since the C
moieties are expelled from the chains at $r=1$, there is only one
critical point for $z_{\text{C}}$.

The average degree of polymerization, $z$, has here two contributions:
one from the chains containing A moieties and one from the isolated
C moieties. The number fraction of isolated C moieties among all chains
is $1-b$, while the fraction of longer chains is $b$. The average
number of moieties on the A containing chains is $2z_{A}+1$, since
A is minority here. Thus, 
\begin{equation}
z=b\left(2z_{\text{A}}+1\right)+\left(1-b\right)\label{eq:z-2}
\end{equation}
\[
=\frac{r+s+1}{r+s-1}.
\]

In order to compute $z_{\text{D}}$, we have to compute the probability
$p_{\text{j}}$ that an inner ``bond'' connecting two A moieties
is of type D. The weight fraction of C moieties that are not an end
is $b^{2}$. Therefore, there is a portion of 
\begin{equation}
p_{\text{j}}=\frac{r}{b^{2}s+r}=\frac{rs}{\left(1-r\right)^{2}+rs}\label{eq:p-1}
\end{equation}
inner bonds that are of type D. Let 
\begin{equation}
t=\frac{1}{z_{\text{A}}}\label{eq:s-1}
\end{equation}
denote the probability per A moiety that a ``chain'' of A moieties
is terminated by an excess C moiety. Again, the same number fraction
distribution applies for the A moieties and the bonds in between leading
to the very same specific degrees of polymerization, see equation
(\ref{eq:z_X}). This number fraction distribution is of most probable
type and given by 
\begin{equation}
n_{\text{N,A}}=\left(1-t\right)^{N-1}t.\label{eq:nND-1}
\end{equation}
The number fraction distribution of the bond sequence between A moieties
where all $N$ of these inner bonds are of type C is therefore 
\begin{equation}
n_{\text{N,C}}=\sum_{N=1}^{\infty}n_{\text{N,A}}(1-p_{\text{i}})^{N}\label{eq:NNC}
\end{equation}
\[
=\frac{t}{1-t}\sum_{N=1}^{\infty}\left[\left(1-t\right)\left(1-p_{\text{i}}\right)\right]^{N}=\frac{t}{1-t}\sum_{N=1}^{\infty}\left[\frac{\left(1-r\right)^{2}}{s}\right]^{N}
\]
\[
=\frac{\left(1-r\right)^{3}\left(s+r-1\right)}{\left(rs+\left(1-r\right)^{2}\right)\left(s-\left(1-r\right)^{2}\right)}.
\]

The remaining portion of $1-n_{\text{NC}}$ of all A containing chains
is the fraction where all D moieties are located. Since the ratio
between inner D and C moieties is $p_{\text{j}}$, we arrive at 
\begin{equation}
z_{\text{D}}=\frac{z_{\text{A}}p_{\text{j}}}{1-n_{\text{N,C}}}=1+\frac{rs}{\left(r+s-1\right)\left(1-r\right)}.\label{eq:zD}
\end{equation}

\begin{figure}
\includegraphics[angle=270,width=1\columnwidth]{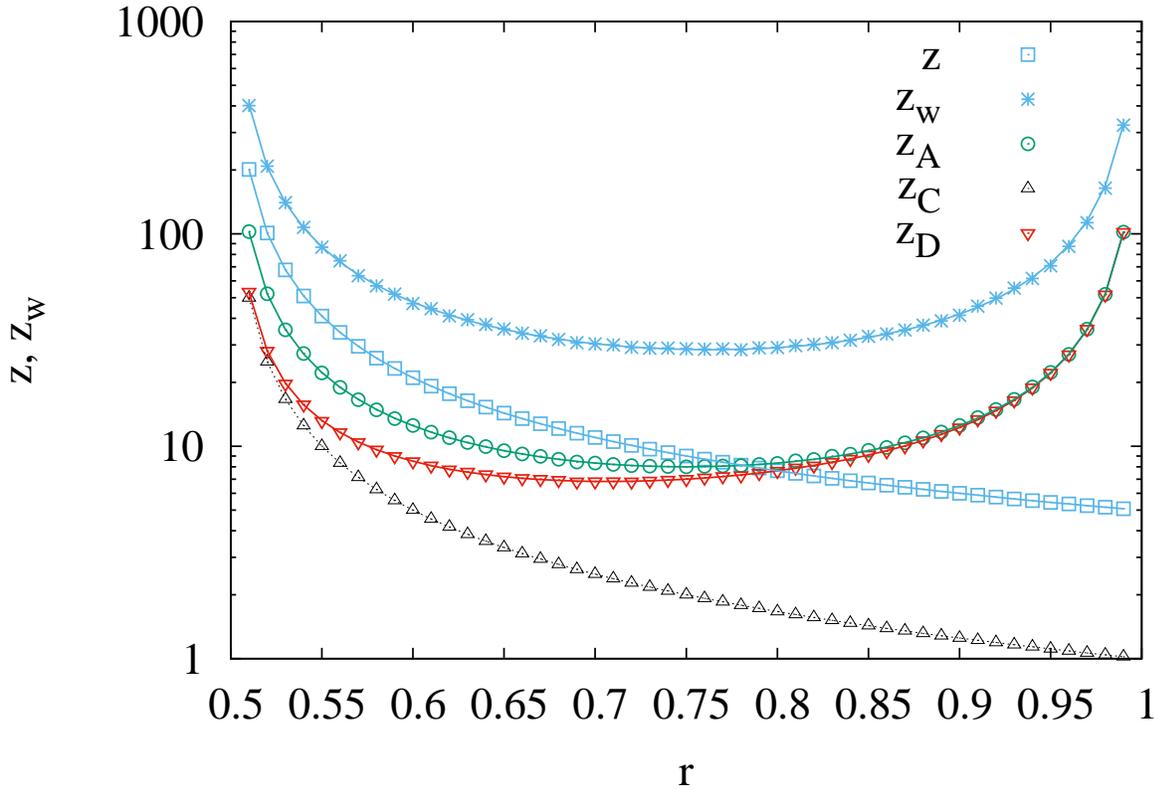}

\caption{\label{fig:Comparison-of-simulation}Comparison of simulation data
of the case $s=1/2$ with theory (lines) for average degree of polymerization
in case of $1/2<r<1$.}
\end{figure}

Obviously, the number average molar mass, $M_{\text{n}}$, is given
by equation (\ref{eq:Mn}), since for $M_{\text{n}}$ only the divergence
at $r+s=1$ defines the two different regimes. The above results are
compared with simulation data in Figure \ref{fig:Comparison-of-simulation}
for the special case of $s=1/2$. The good agreement between both
shows that correlations in the sequence of bonds can be ignored.

For the discussion in the following sections, we also require the
number fraction of stable chains, $n_{\text{N,D}}$, which is given
by 
\begin{equation}
n_{\text{N,D}}=\sum_{N=1}^{\infty}n_{\text{N,A}}p_{\text{i}}^{N}=\label{eq:nnDpi}
\end{equation}

\[
\frac{t}{1-t}\sum_{N=1}^{\infty}r^{N}=\frac{tr}{\left(1-t\right)\left(1-r\right)}.
\]

Similar to the discussion of the preceding sections, we assume most
probable weight distributions for $z_{\text{A}}$, $z_{\text{C}}$,
and $z_{\text{D}}$, which yields 
\begin{equation}
z_{\text{w,A}}=z_{\text{A}}\left(1+b\right)=\frac{s+1-r}{\left(1-r\right)\left(s+r-1\right)},\label{eq:zwa-1}
\end{equation}
\begin{equation}
z_{\text{w,C}}=z_{\text{C}}\left(1+b\right)=\frac{s+1-r}{s+r-1},\label{zwc-1}
\end{equation}
\begin{equation}
z_{\text{w,D}}=z_{\text{D}}\left(1+b\right)=\frac{1-r}{s}+\frac{2r}{s+r-1}+\frac{1}{1-r}.\label{eq:zwd-1}
\end{equation}

As above, computation $M_{\text{w}}$ and $z_{\text{w}}$ requires
to consider the weight distribution of the combined chains, here made
of A, C, and D moieties. As we shall see below, there is typically
$z\gg1$, which allows to estimate $M_{\text{w}}$ in a simplified
manner through 
\begin{equation}
M_{\text{w}}\approx z_{\text{w}}\overline{M}\label{eq:Mwz}
\end{equation}
since any correlation that might be important for small $z$ affects
only a small weight fraction of all chains. Therefore, we discuss
here only the derivation of $z_{\text{w}}.$

There are $2N+1$ moieties per chain that contains $N$ moieties of
type A. The weight fraction of moieties on chains with $N$ moieties
of type A among all A containing chains, $w_{\text{N,A}}$ is derived
using the number fraction distribution of A $N$-mers, $n_{\text{N,A}}$,
and the corresponding average number of moieties $2z_{\text{A}}+1$
on these chains: 
\begin{equation}
w_{\text{N,A}}=\frac{2N+1}{2z_{\text{A}}+1}n_{\text{N,A}}\label{eq:wND-1}
\end{equation}
\[
=\frac{2N+1}{2+t}\left(1-t\right)^{N-1}t^{2}.
\]
The weight average number of moieties on all A containing chains is
then

\begin{equation}
z_{\text{w,A*}}=\sum_{N=1}^{\infty}w_{\text{N,A}}\left(2N+1\right)\label{eq:zw-1}
\end{equation}
\[
=\frac{t^{2}}{2+t}\sum_{N=1}^{\infty}\left(2N+1\right)^{2}\left(1-t\right)^{N-1}
\]
\[
=\frac{t^{2}+8}{\left(2+t\right)t}.
\]

\begin{figure}
\includegraphics[angle=270,width=1\columnwidth]{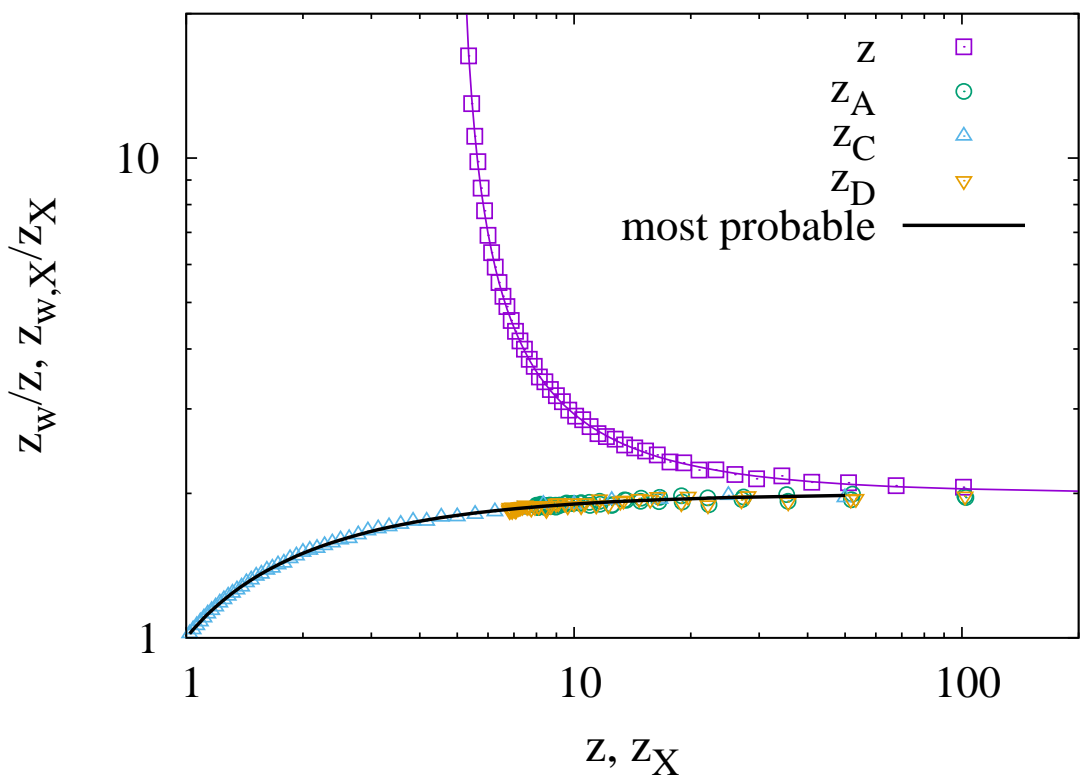}

\caption{\label{fig:Polydispersity-of-different-1}Polydispersity of different
average degrees of polymerization of the specific case of Ref. \citep{Suckow}
compared with theory equation (\ref{eq:zwz-1}) and (\ref{eq:poly Z})
for average molecular weights in case of $1/2<r<1$.}
\end{figure}

In order to obtain $z_{\text{w}}$ of the full sample, we have to
add to $z_{\text{w,A*}}$ the contribution of the weight fraction
$w_{1}$ of non-reacted C moieties among all moieties within the sample.
A weight fraction of $\left(1-b\right)^{2}$ of all C moieties are
non-reacted C moieties, while a number fraction of $s/\left(1+s+r\right)$
is the portion of C moieties among all moieties. Thus, the weight
fraction of non-reacted C moieties, $w_{1}$, among all moieties (putting
$M_{\text{A}}=M_{\text{C}}=M_{\text{D}}=1$) is 
\begin{equation}
w_{1}=\frac{s\left(1-b\right)^{2}}{1+s+r}=\frac{\left(r+s-1\right)^{2}}{s\left(r+s+1\right)}\label{eq:WC-1}
\end{equation}
\textbf{ }thus, 
\begin{equation}
z_{\text{w}}=\left(1-w_{1}\right)z_{\text{w,A*}}+w_{1}=\label{eq:Mw-1}
\end{equation}
\[
=1+\frac{4}{r+s-1}+\frac{2\left(3r+1\right)}{\left(1-r\right)\left(1+r+s\right)}.
\]
This leads to a polydispersity of 
\begin{equation}
\frac{z_{\text{w}}}{z}=\frac{\left(r-1\right)s^{2}+2\left(r-3\right)\left(r+1\right)s+\left(r-1\right)^{3}}{\left(r-1\right)\left(r+s+1\right)},\label{eq:zwz-1}
\end{equation}
which diverges in the limit of $r\rightarrow1$ for reasons as discussed
for the case of $r>1$. The simulation data for the polydispersity
of the special case $s=1/2$ are compared with the corresponding theoretical
expressions in Figure \ref{fig:Polydispersity-of-different-1}.

\section{Discussion\label{sec:Discussion}}

\subsection{\label{sec:Dependence-on-reaction}Dependence on Reaction Rates}

\begin{figure}
\includegraphics[angle=270,width=1\columnwidth]{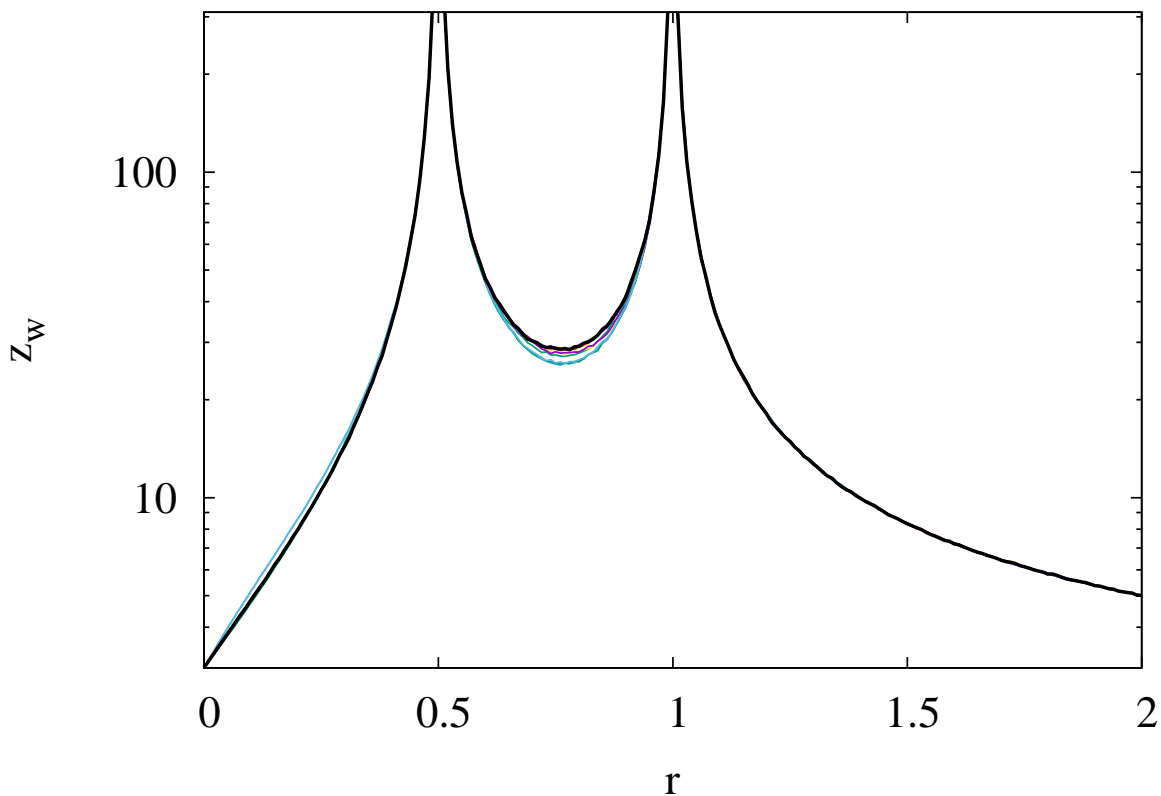}

\caption{\label{fig:Variation-of-the}Variation of the $z_{w}$ upon changing
$k_{i}$, $i=1,...,g$ by one order of magnitude either up or down.
Reference case of equal rates is indicated by the thick black line.
Since the differences are rather small, no legend for all cases studied
was included.}
\end{figure}

In this section, we discuss briefly the impact of reaction rates on
average degrees of polymerization. For this purpose, we ran simulations
where always one of the reaction rates of the reaction scheme shown
in Figure \ref{fig:Scheme of reactions} was either increased or decreased
by one order of magnitude in order to test the sensitivity of the
molecular weight distributions on reaction rates. As pointed out in
the preceding sections, higher order averages like $z_{\text{w}}$
are more sensitive to deviations from the equireactive case. However,
our simulation data for $z_{\text{w}}$ shown in Figure \ref{fig:Variation-of-the}
indicate only a weak dependence on reaction rates that is visible
in the plot only for the two regimes at $r<1$. We have to point out,
that most of the impact of reaction rates stems for the system of
Ref. \citep{Suckow} from the asymmetry of A moieties and is absent
for symmetric moieties that are all unconnected at the onset of the
reactions. In consequence of these observations, we decided not to
discuss full analytical solutions for asymmetric A moieties and omit
also an explicit discussion based upon reaction rates.

\subsection{Post Tuning of the Reactions}

A particularly interesting point of the reactions discussed in the
preceding sections is that the bonds with C moieties behave similar
to reversible bonds (are able to undergo exchange reactions). Thus,
later addition or extraction of C moieties can be used to couple or
uncouple stable chain section that are made entirely of A and D moieties,
since bonds between the latter are stable. The distribution of these
stable sections is the distribution used to compute $n_{\text{N,D}}$
and can be found in each section, see equation (\ref{eq:nND}), (\ref{eq:nND-2}),
and (\ref{eq:nnDpi}). However, one has to be aware that there are
limitations given by the boundaries between the regimes. Thus, the
maximum amount of C moieties that can be taken out in the C-dominated
regime is limited by the condition $r+s>1$ (for homogeneous systems).
Also, addition of C moieties can drive the system from the A-dominated
regime at $r+s<1$ into the C dominated regime with $r+s>1$, however,
the way back is impossible as the total number of bonds to C moieties
cannot be reduced.

Also, the stable parts can be modified by either adding some more
A or D moieties. If minority species are added such these remain minority,
it is sufficient to compute the new ratios $r$ and $s$ after the
modifications and to use these in the corresponding section. If majority
species are added, this can be treated along the D terminated case,
since the addition leads to additional non-reacted moieties of the
majority species. If the addition drives the system across one of
the critical points, the resulting molecular weight distributions
will depend on the particular way how the species are mixed and no
simple predictions are possible. 

Finally, we have to remark that the reversible nature of the exchange
reactions can be ceased and the weight distribution can be fixed by
terminating all unsaturated C groups by the addition of, for instance,
mono-functional $A$ moieties.

\subsection{Optimum Conditions For Minimum Impact of Composition Fluctuations}

Another possibility to utilize this particular competing co-polycondensation
scheme is to run experiments at the minimum of a particular $z_{\text{X}}$
or $z_{\text{w}}$ in the regime with $r<1<r+s$, since there, the
specific degrees of polymerization for $X=A,D$ are rather insensitive
towards fluctuations of $r$ around an average composition, see Figure
\ref{fig:Polydispersity-of-different-1}. Note that we focus here
on the irreversible part of the reactions as composition fluctuations
of the reversible part can be removed by continuous mixing. The minima
are located in the regime $r<1<r+s$ and found for constant $s>0$
through the condition $\partial z_{X}/\partial r=0$ with $X=A,D$.

For $z_{A}$ (see equation (\ref{eq:z_AB-1})), we obtain for the
minimum that
\begin{equation}
s=2-2r.\label{eq:s-3}
\end{equation}
Inserting this $s$ into the expression for $z_{\text{A}}$, equation
(\ref{eq:z_AB-1}), leads to 
\begin{equation}
z_{\text{A}}=\frac{2}{1-r}.\label{eq:z_A}
\end{equation}
Thus, if a desired $z_{\text{A}}$ with low polydispersity should
be obtained, one solves equation (\ref{eq:z_A}) for $r$, which provides
the optimum conditions for synthesis: 
\begin{equation}
r_{\text{opt,A}}=1-\frac{2}{z_{\text{A}}}\label{eq:r-3}
\end{equation}
\begin{equation}
s_{\text{opt,A}}=\frac{4}{z_{\text{A}}}.\label{eq:s-4}
\end{equation}
In similar manner, one finds the optimum conditions for polymerizing
D moieties to a desired $z_{\text{D}}$ that 
\begin{equation}
r_{\text{opt,D}}=1-\frac{2}{z_{\text{D}}}\label{eq:r4}
\end{equation}
\begin{equation}
s_{\text{opt,D}}=\frac{4}{z_{\text{D}}}\left(1-\frac{1}{z_{\text{D}}}\right).\label{eq:s4}
\end{equation}

The sample average degree of polymerization, $z$, has no minimum
for $r<1<r+s$ that could be used to suppress the effect of composition
fluctuations on polydispersity. Thus, if only $z$ needs to be obtained
at minimum polydispersity, one simply replaces all irreversible bonds
by reversible ones such that all composition fluctuations can decay
with time. Finally, $z_{\text{C}}$ has a trivial minimum at $s\rightarrow\infty$
for polydispersity that cannot be used to tune $z_{\text{C}}$. 

Recent work showed that polydispersity may be reduced for polycondensations
that are carried out in small droplets \citep{Szymanski}. However,
we have to remark that such a strategy is not viable for alternating
polycondensations as the relaxation of composition fluctuations is
hindered by the boundaries of these small volumes, which enhances
the polydispersity.

\section{Summary}

In the present paper we have discussed in detail a co-polycondensation,
where one ``dominating'' reacting unit D replaces bonds of the ``non-dominant''
competitor C with the alternating partner A. This model case is representative
for irreversible reactions that compete with exchange reactions for
reactions with the same alternating partner. This process leads to
three distinct regimes that are characterized by the reaction partner
that terminates the polymeric species resulting from the reactions.

For A terminated chains, the competition between D and C is unimportant
as the sum of both is in the minority. Therefore, the reaction as
a whole is qualitatively equivalent to an alternating co-polycondensation
while the specific degrees of polymerization considering only C or
D units are obtained by splitting the A terminated chains into populations
that contain at least one C or D unit respectively. All averages (number,
weight, and specific ones) diverge at the point where the concentrations
of both competitors equal the concentration of the alternating partner,
$[C]+[D]=[A]$.

For D terminated chains, reactions of C units are completely replaced
by reactions with D units such that all C units have no bonds. Here,
the chains are made of A and D units only whereby these react according
to a perfectly alternating polycondensation. Therefore, the corresponding
specific degrees of polymerization of A and D units diverge at $[D]=[A]$,
which holds also for the corresponding weight averages. On the contrary,
the large number of C units enforces a finite number average molar
mass.

For the C terminated regime, the C units are partially replaced by
D units inside the chains, such that the specific degree of polymerization
of C units diverges at the boundary to the A terminated regime, while
it becomes unity at the boundary to the D terminated regime. The specific
degrees of polymerization of D and A units diverge both at both boundaries
and show a distinct minimum in between, that provides optimum conditions
where the reactions are least sensitive towards composition fluctuations.
For a desired specific degree of polymerization, one can adjust the
stoichiometric ratios such that reactions occur at this least sensitive
point. This can be used to reduce the polydispersity of the resulting
polymers or to design new co-polycondensation strategies where high
molar mass products have been unavailable previously because of incomplete
mixing. All weight averages except of C units diverge at both boundaries
to the A and D dominated regimes, while the number average molar mass
diverges only at the boundary to the A terminated regime.

Exact theoretical expressions for all number and weight average degrees
of polymerizations and most molar masses were provided for homogeneous
model reactions in the equireactive case (all reaction rates are identical)
and at the absence of cyclization. These analytical results were double
checked by Monte-Carlo simulations that model the reaction scheme.
Furthermore, the kinetic equations describing all reactions were integrated
numerically to compare with the simulation data. In all cases, perfect
agreement between simulations and kinetic equations was obtained.

Beyond the general discussion of such reactions, we also compared
with the specific situation of Ref. \citep{Suckow}, where the initial
conditions and the asymmetry of the A units introduce a non-random
bond sequence and a shift in the specific degree of polymerization
of the C units. These resulting correlations become less important
the more additional reactions occur as introduced here by D units,
such that these can be ignored once there were sufficient D units
added to arrive at $[C]+[D]>[A]$. While non-random sequences and
reaction rates are important for weight averages and the specific
degree of polymerization of the competing units, these correlations
have no impact on the average molar mass and the specific degree of
polymerization of the alternating partner.

\section{Appendix: Ideal Alternating Polycondensation}

The general mean field solution for alternating polycondensation of
molecules of arbitrary functionality was given first by Stockmayer
\citep{Stockmayer} in 1952. Using Flory's simplifying assumptions
of no cyclization and equal reactivity of all reactive groups, the
number average, $N_{\text{n}}$, and the weight average degrees of
polymerization, $N_{\text{w}}$, as well as $M_{\text{n}}$ and $M_{\text{w}}$
were computed exactly. At full conversion of the minority species,
which is typically reached given enough time for reactions and a stoichiometry
different from one, $r\ne1$, the classical result for the degree
of polymerization reduces to a simple expression that is solely a
function of $r$ for two functional units. The original results were
provided in Ref. \citep{Stockmayer} without derivation. We provide
below a quick derivation for the ideal case of a strictly alternating
linear polycondensation.

Consider a linear polymerization of two di-functional monomers $X$
and $Y$ that undergo only reactions between $X$ and $Y$ groups
without side reactions. Let $r$ denote the ratio of the concentrations
$[X]$ and $[Y]$ of reactive groups on $X$ molecules and $Y$ molecules,
respectively, 
\begin{equation}
r=\frac{[X]}{[Y]}.\label{eq:r1}
\end{equation}
For $r\ne1$, we assume that the reaction terminates when all reactive
groups of the minority species are consumed. Therefore, for strictly
bimolecular recombination without side reactions, the ratio $r$ connects
also the conversions $p_{\text{X}}$ and $p_{\text{Y}}$ of $X$ and
$Y$ groups at complete reactions: 
\begin{equation}
r=\frac{p_{\text{Y}}}{p_{\text{X}}}.\label{eq:r}
\end{equation}
Due to the symmetry of the problem, it is sufficient to discuss only
the case of $r<1$ . The end of the reactions is characterized by
the minority conversion of here $p_{\text{X}}=1$, which yields 
\begin{equation}
r=p_{\text{Y}}.\label{eq:r2}
\end{equation}
The non-reacted groups of the majority $Y$ species terminate the
chains. If there is no cyclization, all chains will be linear, will
have an odd degree of polymerization because of the alternating sequence
of $X$ and $Y$, and all chain ends must be of majority type $Y$.
The number fraction of $1-p_{\text{Y}}$ of non-reacted groups (``chain
ends'') among all majority $Y$ groups is used to compute the average
number of 
\begin{equation}
N_{\text{Y}}=\frac{1}{1-p_{\text{Y}}}=\frac{1}{1-r}\label{eq:Nb}
\end{equation}
majority monomers of type $Y$ per molecule. Since there is always
one more majority monomer than minority monomer per linear chain,
we obtain for the minority species 
\begin{equation}
N_{\text{X}}=N_{\text{Y}}-1=\frac{r}{1-r}\label{eq:Na}
\end{equation}
$X$ monomers per chain. The average number of monomers per chain
of both types $A$ and $B$ defines the number average degree of polymerization
of the linear chains, 
\begin{equation}
N_{\text{n}}=N_{\text{X}}+N_{\text{Y}}=\frac{1+r}{1-r}.\label{eq:N_n}
\end{equation}
The reaction is of poly-condensation type, with a most probable number
fraction distribution $n_{\text{N}_{\text{Y}}}$ that is controlled
by the number $N_{\text{Y}}$ of majority monomers $Y$ per chain.
The minority monomers $X$ act effectively as bonds between $Y$ monomers.
Thus, the stoichiometric ratio is equivalent to the portion of existing
bonds for $r<1$, see equation (\ref{eq:r2}). This allows to rewrite
the expression for the number fractions of chains of a most probable
distribution (see equation 1.52 or Ref. \citep{Rubinstein}) by using
$N_{\text{Y}}$ as the degree of polymerization and $r$ as the conversion:
\begin{equation}
n_{N_{\text{Y}}}(r)=(1-r)r^{N_{\text{Y}}-1}.\label{eq:NBn}
\end{equation}

The weight average degree of polymerization of the linear chains,
$N_{\text{w}}$, is computed in the standard way by considering the
first and second moment, $m_{1}$ and $m_{2}$, of this distribution:
\begin{equation}
N_{\text{w}}=\frac{m_{2}}{m_{1}}.\label{eq:N_w}
\end{equation}
The first moment is 
\begin{equation}
m_{1}=\sum_{N_{\text{Y}}=1}^{\infty}n_{N_{\text{Y}}}(r)(2N_{\text{Y}}-1)=\frac{1+r}{1-r}=N_{\text{n}},\label{eq:m1-1}
\end{equation}
where $2N_{\text{Y}}-1$ is the degree of polymerization of a chain
counting both $X$ and $Y$ monomers. Similarly, one obtains for the
second moment 
\begin{equation}
m_{2}=\sum_{N_{\text{B}}=1}^{\infty}n_{N_{\text{B}}}(r)(2N_{\text{B}}-1)^{2}=\frac{1+6r+r^{2}}{\left(1-r\right)^{2}}.\label{eq:m2-1}
\end{equation}
Thus, 
\begin{equation}
N_{\text{w}}=\frac{1+6r+r^{2}}{1-r^{2}}.\label{eq:N_w1}
\end{equation}
These results agree with other mean field derivations, see for instance
Refs. \citep{Stockmayer,Macosko}. The poly-dispersity of this reaction
is 
\begin{equation}
\frac{N_{\text{w}}}{N_{\text{n}}}=\frac{1+6r+r^{2}}{\left(1+r\right)^{2}}=1+\frac{4r}{\left(1+r\right)^{2}}.\label{eq:PDI}
\end{equation}
As mentioned in the introduction, we note that both $N_{\text{n}}$
and $N_{\text{w}}$ diverge for $r=1$, while the polydispersity remains
finite and converges to $2$ for $r\rightarrow1$ similar to the classical
case of a polycondensation of the same monomers, but with a different
function.

For the above ideal alternating polycondensation, each molecules contains
at least one Y moiety. Thus, 
\begin{equation}
z_{\text{Y}}=N_{\text{Y}}.\label{eq:z_Y}
\end{equation}
The corresponding result for the minority species is computed by re-normalizing
$N_{\text{X}}$ to the portion of molecules that contain $X$ monomers,
which is here the portion of chains that are no $Y$ monomers, 
\begin{equation}
z_{\text{X}}=\frac{N_{\text{X}}}{1-n_{N_{\text{Y}}=1}(r)}=\frac{N_{\text{X}}}{1-\left(1-r\right)}=\frac{1}{1-r}=z_{\text{Y}}.\label{eq:z_X}
\end{equation}
Thus, both majority and minority species share the same specific degree
of polymerization. These specific weight averages are characterized
by the same number fraction distribution, equation (\ref{eq:NBn}),
which is of most probable type. Let $z_{\text{w,X}}$ denote the weight
average of the specific degree of polymerization of species $X$.
For a most probable distribution, see equation (\ref{eq:NBn}), it
is known that polydispersity scales as \citep{Rubinstein} 
\begin{equation}
\frac{z_{\text{w,X}}}{z_{\text{X}}}=1+r,\label{eq:most probable}
\end{equation}
since $r$ refers here to the fraction of existing bonds, i.e. the
conversion. Therefore, 
\begin{equation}
z_{\text{w,X}}=\left(1+r\right)z_{\text{X}}=\frac{1+r}{1-r}=z_{\text{w,Y}}=N_{\text{n}}.\label{eq:poly Z}
\end{equation}

\section*{Acknowledgement}

M. Lang thanks the Deutsche Forschungsgemeinschaft (DFG) for funding
LA2735/5-1.

\section*{Conflict of interest}

The authors declare no conflicts of interest.

\newpage

\section*{Table of Contents}

\includegraphics[angle=270,width=1\columnwidth]{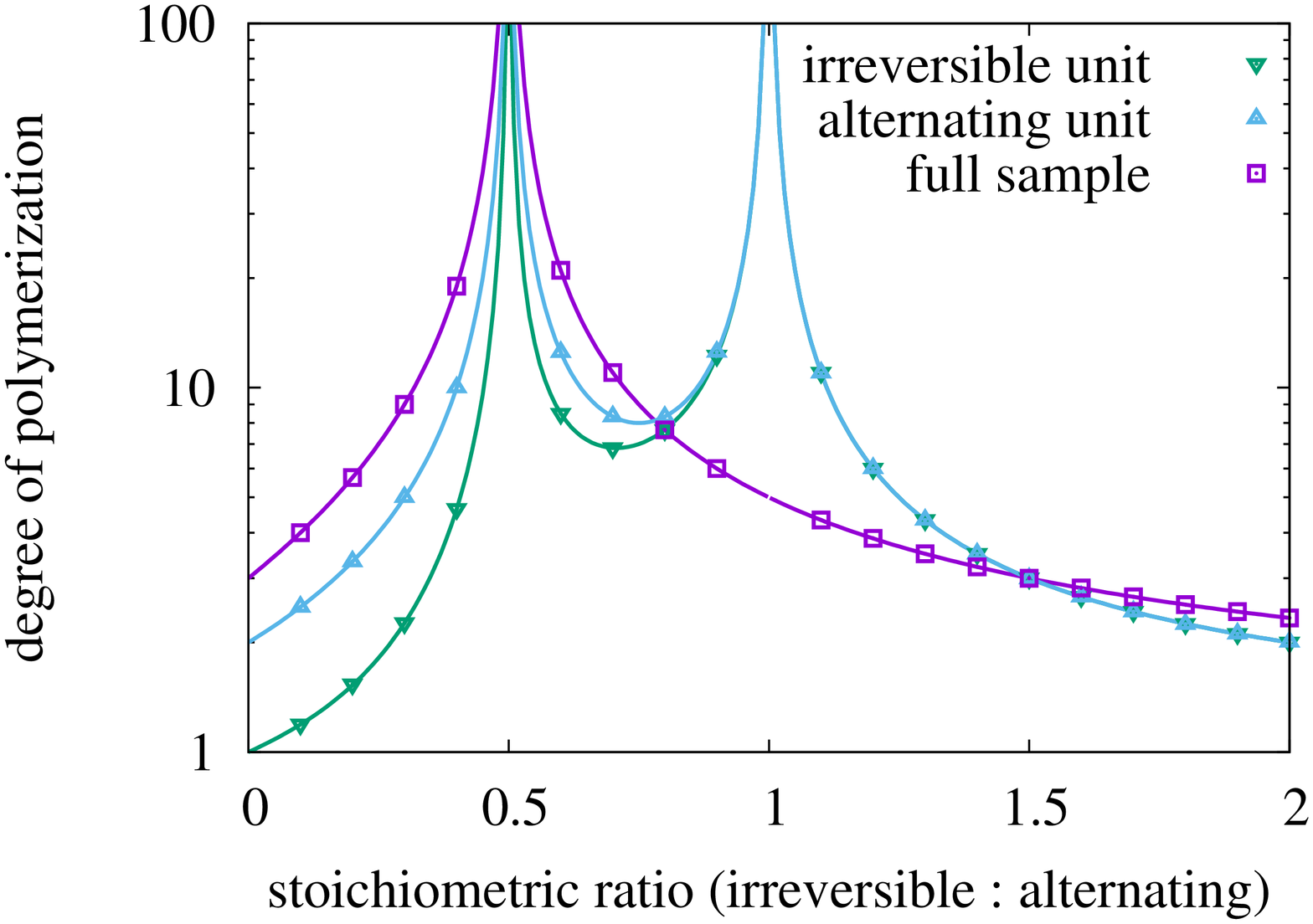}

\bigskip\bigskip

Two reactive units compete for the linear co-polycondensation with an alternating third unit such that bonds of the irreversibly reacting unit replace bonds which undergo exchange reactions. Two peaks for the degree of polymerization are observed for the dominatingirreversibly reacting and alternating unit as a function of composition. The first peak and the  plateau between the peaks can be tuned by adjusting the concentration of units that undergo exchange reactions.

Two reactive units compete for the linear co-polycondensation with
an alternating third unit such that irreversible bonds replace bonds
which undergo exchange reactions. Two peaks for the degree of polymerization
are observed for the dominating and alternating unit as a function
of composition. The first peak and the range between the peakscan
be tuned by adjusting the concentration of the units that undergo
exchange reactions.
\end{document}